\documentclass[journal]{IEEEtran}

\usepackage{booktabs,threeparttable}
\usepackage{graphics} % for pdf, bitmapped graphics files
\usepackage{epsfig} % for postscript graphics files
\usepackage{mathptmx} % assumes new font selection scheme installed
\usepackage{times} % assumes new font selection scheme installed
\usepackage[fleqn]{amsmath} % assumes amsmath package installed
\usepackage{amssymb}  % assumes amsmath package installed
\usepackage{amsthm}
\usepackage{fancyhdr}
\usepackage{subfigure}
\usepackage{graphicx}
\usepackage{algorithmic}
\usepackage{algorithm}
\usepackage{mathrsfs}
\usepackage[]{units}
\newcommand{\makenumbered}[2]{
\newcounter{#1}[section]
\newenvironment{#1}%
{\begin{list}{}{
  \setlength{\itemsep}{0in}%
  \setlength{\labelwidth}{0in}%
  \setlength{\labelsep}{0in}%
  \setlength{\rightmargin}{0in}%
  \setlength{\leftmargin}{0in}%
  \setlength{\parsep}{0in}%
  \setlength{\listparindent}{\parindent}}%
  \refstepcounter{#1}%
  \item[] {\small\bf #2} }%
{\end{list}}}
\makenumbered{theoremn}{Theorem \thetheoremn:}\def\thetheoremn{\thesection.\arabic{theoremn}}
\makenumbered{lemman}{Lemma \thelemman:}\def\thelemman{\thesection.\arabic{lemman}}

\begin{document}

\title{Efficient Channel-Hopping Rendezvous Algorithm Based on Available Channel Set}

\author{Lu Yu, Hai Liu,~\IEEEmembership{Member,~IEEE}, Yiu-Wing Leung,~\IEEEmembership{Senior Member,~IEEE}, \\Xiaowen Chu,~\IEEEmembership{Senior Member,~IEEE}, and~Zhiyong~Lin
        % <-this % stops a space
\IEEEcompsocitemizethanks{\IEEEcompsocthanksitem L. Yu, H. Liu, Y.-W. Leung, and X. Chu are with the Department of Computer Science, Hong Kong Baptist University, Kowloon Tong, Hong Kong SAR.\protect\\
E-mail: \{lyu, hliu, ywleung, chxw\}@comp.hkbu.edu.hk.
\IEEEcompsocthanksitem Z. Lin is with the Department of Computer Science, GuangDong Polytechnic Normal University, GuangZhou, China.\protect\\ Email: sophyca.lin@gmail.com.
\IEEEcompsocthanksitem A preliminary version of this paper was presented at ICC2014.}% <-this % stops a space
\thanks{}}

\maketitle

\begin{abstract}
In cognitive radio networks, rendezvous is a fundamental operation by which two cognitive users establish a communication link on a commonly-available channel for communications. Some existing rendezvous algorithms can guarantee that rendezvous can be completed within finite time and they generate channel-hopping (CH) sequences based on the whole channel set. However, some channels may not be available (e.g., they are being used by the licensed users) and these existing algorithms would randomly replace the unavailable channels in the CH sequence. This random replacement is not effective, especially when the number of unavailable channels is large. In this paper, we design a new rendezvous algorithm that attempts rendezvous on the available channels only for faster rendezvous. This new algorithm, called \emph{Interleaved Sequences based on Available Channel set} (ISAC), constructs an odd subsequence and an even subsequence and interleaves these two subsequences to compose a CH sequence. We prove that ISAC provides guaranteed rendezvous (i.e., rendezvous can be achieved within finite time). We derive the upper bound on the maximum time-to-rendezvous (MTTR) to be $O(m)$ $(m\leq Q)$ under the symmetric model and $O(mn)$ $(n\leq Q)$ under the asymmetric model, where $m$ and $n$ are the number of available channels of two users and  $Q$ is the total number of channels (i.e., all potentially available channels). We conduct extensive computer simulation to demonstrate that ISAC gives significantly smaller MTTR than the existing algorithms.
\end{abstract}


\section{Introduction}
\IEEEPARstart{I}{n} conventional wireless networks, a significant portion of the licensed spectrum is under-utilized while the unlicensed spectrum is over-crowded [1]. Cognitive radio networks (CRNs) can flexibly exploit the spectrum to improve its utilization. In CRNs, each unlicensed user (or cognitive user, secondary user, SU) uses a cognitive radio to: i) identify the vacant portions of the licensed spectrum by applying a spectrum sensing method [2], and ii) use a vacant portion for communications by adjusting its transmission parameters. For ease of operation, the spectrum is typically divided into channels. If two or more cognitive users want to communicate with each other, they must select a channel which is available to all of them and establish a communications link on this channel. This operation is known as \emph{rendezvous}. For convenience, we refer to ``cognitive users" as ``users" in this paper.
\par Channel hopping (CH) is one of the most representative approaches to rendezvous. Using this approach, each cognitive user follows a certain CH sequence to hop among the channels. When the cognitive users hop on the same channel at the same time, they can exchange handshaking messages to establish a communication link on this channel. Several rendezvous algorithms have been proposed in the literature for generating CH sequences (a detailed survey can be found in [4] and a brief review is given in Section II). They can be divided into two types: i) guaranteed rendezvous (i.e., rendezvous can be completed within finite time) [5, 8, 23], and ii) unguaranteed rendezvous (i.e., rendezvous may take infinite time) [9, 10]. The former type is desirable because it can provide better perception to the users.

\par For the existing algorithms that provide guaranteed rendezvous [5, 8, 23], they generate the CH sequences based on the whole channel set (i.e., set of all channels). In practice, some channels may not be available because: i) these channels are being used by the licensed users, or ii) the channels are not being used but the cognitive users cannot identify these idle channels because of the practical limitations of the spectrum sensing methods [3]. In fact, the available channel set is usually a small subset of the whole channel set. For example, work in [3] optimizes the detection (sensing) time for channel efficiency. It shows that when the SNR (signal-to-noise ratio) is $-3$dB, the SUs can only detect one idle channel in 15\% of the total idle period. In other words, on average, at most 15\% of channels can be correctly detected as ``available" for data transmission of SUs. If a cognitive user follows a CH sequence which includes both the available and the unavailable channels, this user would waste time in attempting rendezvous on the unavailable channels. This would increase the time to complete rendezvous, especially when the available channel set is a small subset of the whole channel set. Table I summarizes the characteristics of existing representative CH algorithms.
\begin{table*}
\caption{Summary of Representative CH Algorithms}
\centering
\newcommand{\tabincell}[2]{\begin{tabular}{@{}#1@{}}#2\end{tabular}}
\begin{tabular}{|c|c|c|c|c|c|}
\hline
Algorithms&\tabincell{c}{Channel Set based on\\(Available or Whole)}&Guaranteed rendezvous&\tabincell{c}{Upper bounds on MTTR\\under the symmetric model}&\tabincell{c}{Upper bounds on MTTR \\under the asymmetric model} \\
\hline
ISAC&Available Channel Set&YES&$2m_p-1$&$2nm_p-G+1$\\
\hline
MC/MMC [9]&Available Channel Set&NO&Infinity&Infinity\\
\hline
Random [10]&Available Channel Set&NO&Infinity&Infinity\\
\hline
Jump-Stay [5]&Whole Channel Set&YES&$3P$&$6QP(P-G)$\\
\hline
Enhanced Jump-Stay [7]&Whole Channel Set&YES&$4P$&$4P(P+1-G)$\\
\hline
CRSEQ [8]&Whole Channel Set&YES&$\geq 3P^2-4P+1$ and $\leq P(3P-1)$&$P(3P-1)$\\
\hline
ARMMC [27]&Whole Channel Set&NO&Infinity&Infinity\\
\hline
\end{tabular}
\begin{tablenotes}
\footnotesize
\item[a] Remarks: $m$, $n$ are the numbers of available channels of two users, respectively; $m_p$ is the smallest prime number which is not smaller than $m$; $Q$ is the total number of channels (i.e. all potentially available channels); $P$ is the smallest prime number which is not smaller than $Q$; $G$ is the number of commonly-available channels of the two users.
\end{tablenotes}
\end{table*}

\par If the cognitive users attempt rendezvous on the available channels only, the time to complete rendezvous can be reduced. The following existing algorithms exploit this basic observation but they cannot guarantee that rendezvous can be completed within finite time:
\begin{enumerate}
\item The random algorithm [10] randomly selects an available channel in each time slot and attempts rendezvous on this channel in this time slot. It cannot provide guaranteed rendezvous.
\item For any existing rendezvous algorithm that generates CH sequences based on the whole channel set, after it has generated a CH sequence, a random replacement operation is executed to randomly replace the unavailable channels in the CH sequence by the available ones [5-6]. However, random replacement is less effective than carefully designed sequences. In fact, when the available channel set is a very small subset of the whole channel set, the rendezvous algorithm with random replacement would degrade to a random algorithm.
\item Two algorithms proposed in [9] (namely, modular clock algorithm and modified modular clock algorithm) generate CH sequences based on the available channel set. However, these algorithms do not guarantee that rendezvous can be completed within finite time.
\end{enumerate}
\par In this paper, we propose a new rendezvous algorithm, called \emph{interleaved sequences based on available channel set} (ISAC), for CRNs. Our contributions are three-fold.
\begin{enumerate}
\item \emph{Algorithm design:} We design ISAC to realize two desirable properties: i) CH sequences are generated based on the available channel set for better rendezvous performance, and ii) rendezvous can certainly be completed within finite time.
\item \emph{Algorithm analysis:} We prove that ISAC provides guaranteed rendezvous. In particular, we derive upper bounds on the maximum time-to-rendezvous (MTTR) for two popular models of channel availability (namely, symmetric model and asymmetric model [7, 11, 12, 13] and these models will be defined in Section II). These upper bounds are expressed in terms of the number of available channels instead of the total number of channels. Table I shows the upper bounds on MTTR of the existing algorithms, especially when the available channel set is a small subset of the whole channel set.
\item \emph{Algorithm evaluation:} We conduct extensive simulation for performance evaluation and demonstrate that ISAC gives significantly smaller MTTR than the existing algorithms. ISAC is suitable to many QoS-concern applications of CRNs where communications are required to be realized within a specific duration (i.e., small MTTR is desired).
\end{enumerate}
\par The rest of this paper is organized as follows. We review the related work in Section II, present the system model and problem definition in Section III, design the ISAC algorithm and theoretically analyze its properties in Section IV, present simulation results for performance evaluation in Section V, and conclude our work in Section VI.

\section{Related Work}
\par In the literature, there are two models to describe the channel availability [7, 11, 12, 13]:
\begin{itemize}
\item \emph{Symmetric Model:} In this model, all cognitive users have the same channel availability. This model is suitable when the cognitive users are close to each other.
\item \emph{Asymmetric Model:} In this model, different cognitive users may have different channel availability. This model is suitable when the cognitive users are located at different positions relative to the licensed users. If some cognitive users want to communicate with each other, they must have at least one commonly available channel.
\end{itemize}
\par The existing CH algorithms can be classified into two categories based on their structures: i) centralized systems where a central server is operated to allocate channels to the cognitive users, and ii) decentralized systems where there is no central server. The decentralized systems can be further classified into two subcategories: i) using a common control channel (CCC), and ii) without using CCC. Fig. 1 shows a possible taxonomy of the existing rendezvous algorithms.
\begin{figure}
\centering
\includegraphics[width=0.5\textwidth]{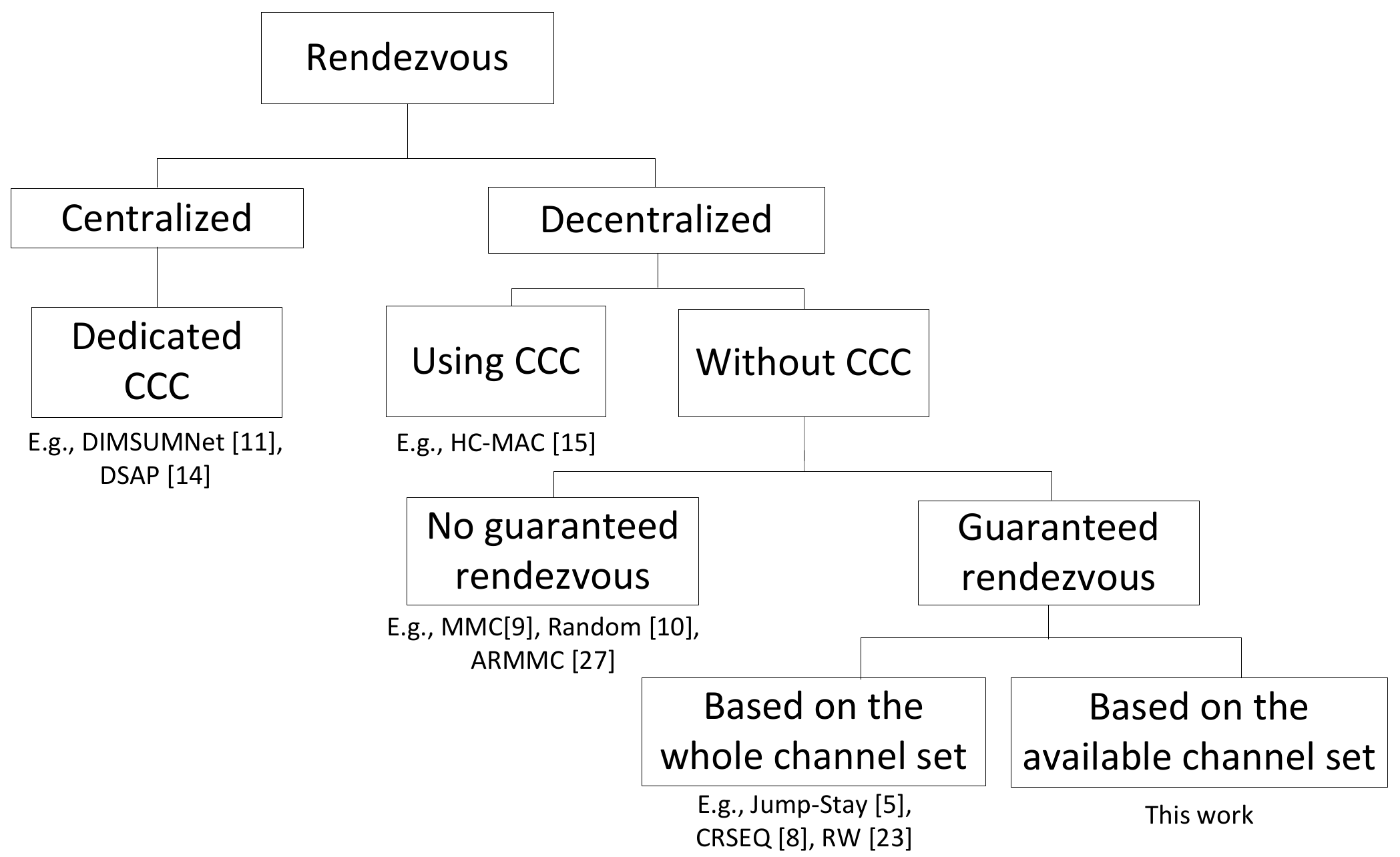}
\caption{A taxonomy of the existing rendezvous algorithms.}
%\vspace{-0.2in}
\end{figure}
\par \emph{Centralized systems}: In centralized systems, such as DIMSUMNet [11] and DSAP [14], a centralized server is operated to schedule the data exchanges among users. With a centralized server for global coordination, this approach eases the rendezvous process but it involves the overhead of maintaining the server and this server is a single point of failure [4].
\par \emph{Decentralized systems using CCC}: In decentralized system, a channel is preselected as a CCC. In [15] and [16], a global CCC is preselected and known to all users. In [17] and [18], a cluster-based control-channel method was proposed, in which a local CCC is selected for each cluster. However, the extra costs in establishing and maintaining the global/local CCCs are considerable.
\par \emph{Decentralized systems without using CCC}: This approach does not use CCC and hence it is known as blind rendezvous [5]. With this desirable feature, this approach has drawn significant attention in the literature and some effective algorithms for blind rendezvous have been proposed (e.g., Jump-Stay [5][6], M-/L-QCH [19] and ASYNC-ETCH [20]). Some algorithms generate CH sequence based on available channels only and others extend the CH hopping sequence into the whole channel set. Based on it we can further divide decentralized systems without using CCC into two types.
\begin{itemize}
    \item\emph{Algorithms based on the whole channel set}: Majority of the existing rendezvous algorithms generate CH sequence based on the whole channel set. Two CH algorithms M-/L-QCH were proposed in [19] based on quorum systems which can guarantee a rendezvous of users under the symmetric model. Another CH algorithm called A-QCH was proposed in [19] for the asynchronous systems. However, A-QCH is only applicable to systems with two channels. Bahl et al. designed a link-layer protocol named SSCH [21]. Each user can select more than one pair of channels and generate the CH sequence based on these pairs. It was designed to increase the capacity of IEEE 802.11 networks. However, similar to M-/L-QCH, they are not applicable under the asymmetric model. Authors in [22] proposed a deterministic approach in which each user is scheduled to broadcast on every channel in an exhaustive manner. In [23], authors presented a ring-walk (RW) algorithm which guarantees the rendezvous under both models. In RW, each channel is represented as a vertex in a ring. Users walk on the ring by visiting the vertices (channels) with different velocities and rendezvous is guaranteed since users with lower velocities will be caught by users with higher velocities. However, RW requires that each user has a unique ID and knows the upper bound of network size. Yang et al. proposed two significant algorithms, namely deterministic rendezvous sequence (DRSEQ) [24] and channel rendezvous sequence (CRSEQ) [8], which provide guaranteed rendezvous for the symmetric model and the asymmetric model, respectively. In CRSEQ, the sequence is constructed based on triangle numbers and modular operations. In terms of MTTR, CRSEQ is quite good under the asymmetric model but it does not perform well under the symmetric model. Bian et al. [25] presented an asynchronous channel hopping (ACH) algorithm which aims to maximize rendezvous diversity. It assumes that each user has a unique ID and ACH sequences are designed based on the user ID. Though the length of user ID is a constant, it may result in a long TTR in practice given that a typical MAC address contains 48 bits. In [26], an efficient rendezvous algorithm based on Disjoint Relaxed Difference Set (DRDS) was proposed while DRDS only costs a linear time to construct. They proposed a distributed asynchronous algorithm that can achieve and guarantee fast rendezvous under both the symmetric and the asymmetric models. They derived the lower bounds of MTTR which are $3P$ and $3P^2+2P$ under the symmetric model and the asymmetric model, respectively. They showed that it is nearly optimal. There are other algorithms in this category such as synchronous QCH [19], SYNCETCH and ASYNC-ETCH [20], AMRCC [27], C-MAC [17], and MtQS-DSrdv [29]. Due to limited space, these algorithms are not reviewed and readers may refer to the survey in [4] for details.
    \item\emph{Algorithms based on available channel set only}: Recently, there are several CH algorithms which generate CH sequence based on available channel set.  Available channel set is a subset of the whole channel set. A notable work by Theis et al. presented two CH algorithms: modular clock algorithm (MC) and its modified version MMC for the symmetric model and the asymmetric model, respectively [9]. The basic idea of MC and MMC is that each user picks a proper prime number and randomly selects a rate less than the prime number. Based on the two parameters, the user generates its CH sequence via pre-defined modulo operations. Although MC and MMC are shown to be effective, both algorithms cannot guarantee the rendezvous if the selected rates or the prime numbers of two users are identical.
 \end{itemize}
 \par To the best of our knowledge, among all existing rendezvous algorithms, only the random algorithm [10] and the MC/MMC algorithms [9] generate CH sequences based on the available channel set; but these existing algorithms cannot guarantee that rendezvous can be completed within finite time (see Table I).
 \section{System Model And Problem Definition}
 \par We consider a CRN consisting of $K$ $(K\geq 2)$ users. The licensed spectrum is divided into $Q$ $(Q\geq 1)$ non-overlapping channels $c_1,c_2,...,c_Q$, where $c_i$ is called channel $i$ $(i=1,2,...,Q)$. Let $C$ be the whole channel set $\{c_1,c_2,...,c_Q\}$. Let $C^i\in C$ be the set of available channels of user $i$ $(i=1,2,...,K)$, where a channel is said to be available to a user if the user can communicate on this channel without causing interference to the licensed users. The available channels can be identified by any spectrum sensing method (e.g., [11]).
 \par Time is divided into slots of equal duration. Time synchronization among the users is not needed (i.e., different users may start their time slots at different time) as long as the handshaking messages for rendezvous can be exchanged in the overlapped interval of two time slots of two respective users [5].
 \par \emph{Rendezvous Problem for two users:} Consider any two users who may have different channel availability and may start the rendezvous operation at different time. The problem is to determine a CH sequence for each user, such that these users will hop on a commonly-available channel in the same time slot within finite time.
  \par \emph{Rendezvous for multiple users in multi-hop networks:} Suppose there are multiple users. They may have different channel availability, they may start the rendezvous operation at different time, or they may be in a multi-hop network. These users can achieve rendezvous as follows: iteratively solve the above rendezvous problem to achieve pairwise rendezvous for two users in order to achieve global rendezvous for all users [5-7]. In this process, the crucial issue is to solve the above rendezvous problem for two users. In the following, we focus on solving the rendezvous problem for two users.
 \section{ISAC Algorithm}
\subsection{Algorithm Design}
\par Consider any two users (say, users 1 and 2). The user who initiates communication assumes the role of \emph{sender} while the other user assumes the role of \emph{receiver}. For the sender (say, user 1), let there be $m$ available channels and the available channel set be $C^1=\{C^1_1, C^1_2, C^1_3,..., C^1_m\}$. For the receiver (user 2), let there be $n$ available channels and the available channel set be $C^2=\{C^2_1, C^2_2, C^2_3,..., C^2_n\}$. Under the symmetric model, $C^1$ is equal to $C^2$; under the asymmetric model, $C^1$ may not be equal to $C^2$. Let $G$ be the number of channels commonly available to the sender and the receiver (i.e., $G$ is equal to the number of elements in $C^1\cap C^2$).
\par Using ISAC, the sender generates a CH sequence based on the available channel set $C^1$ while the receiver generates another CH sequence based on the available channel set $C^2$. We describe how ISAC generates these CH sequences in the following.
\par \textbf{Sender-role Sequence: }Given $C^1=\{C_1^1, C_2^1, ..., C_m^1\}$, $m_p$ be the smallest prime number which is not smaller than $m$. We first expand $C^1$ to get a total of $m_p$ channels and represent these channels in the following multiset (i.e., a set in which the elements may appear more than once [30]):
\centerline{$C^1_*=\{C_1^1, C_2^1, ..., C_m^1, C_{m+1}^1, C_{m+2}^1, ..., C_{m_p}^1\}$}
where $C_h^1$ $(h=m+1, ...,m_p)$ are randomly selected from $\{C_1^1, C_2^1, ..., C_m^1\}$. The sender-role sequence is generated in rounds and each round contains $m_p$ time slots. A starting channel is randomly selected from $C^1_*$. Then, sender moves to the next channel index in $C^1_*$ and keeps hopping on $m_p$ channels in the round-robin fashion. Fig. 2(a) shows the structure of sender-role sequence. Algorithm 1 shows the detailed steps for generating a sender-role sequence. For example, in Fig. 2(b), $m=2$, $m_p=2$, $C^1=\{1, 2\}$ and $C^1_*=\{1, 2\}$. When starting channel is channel 2, the CH sequence is $\{2, 1, 2, 1, ...\}$.
\begin{figure}
\centering
\subfigure[Structure of Sender-role Sequence (starting channel is $C_k^1$)]{
\label{fig:subfig:a} %% label for first subfigure
\includegraphics[scale=0.9]{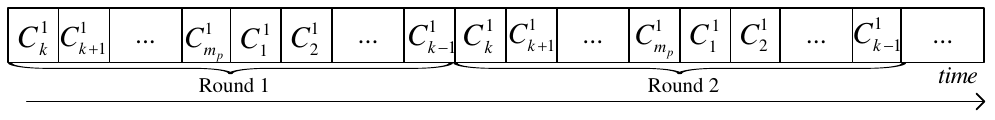}}
\hspace{1in}
\subfigure[Sender-role Sequence when $C^1=\{1, 2\}$, starting channel is channel 2]{
\label{fig:subfig:b} %% label for second subfigure
\includegraphics[scale=0.6]{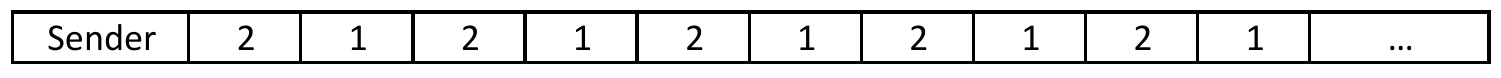}}
\caption{Sender-role Sequence.}
\label{fig:subfig} %% label for entire figure
%\vspace{-0.2in}
\end{figure}
\par \textbf{Receiver-role Sequence: }Given $C^2=\{C_1^2, C_2^2, ..., C_n^2\}$, it randomly generates a permutation of $C^2: \{C_{l_1}^2, C_{l_2}^2, ..., C_{l_n}^2\}$ first. The receiver-role sequence is composed of two interleaving subsequences: Odd Subsequences (OS) and Even Subsequences (ES). That is, the channels in time slot $1, 3, 5, ..., (2n-1)$ follows OS and the channels in time slot $2, 4, 6, ..., (2n)$ follows ES. Fig. 3(a) shows the structure of odd subsequence of receiver-role. $OS_i$ means the $i$-th time slot in OS. OS is generated in rounds and each round contains $n$ time slots where $n$ is the number of channels in the available channel set of the receiver. Receiver keeps hopping on $n$ channels with the order $\{C_{l_1}^2, C_{l_2}^2, ..., C_{l_n}^2\}$ in round-robin fashion alternately. Fig. 3(b) shows the structure of Even Subsequence of receiver-role. ES is generated in rounds and each round contains $n$ time slots. $ES_i$ means the $i$-th time slot in ES. The sequence in the first round is $\{C_{l_1}^2, C_{l_2}^2, ..., C_{l_n}^2\}$. In the next round, all channel indices are left-shifted by 1 to generate a new permutation of $C^2$. That is $\{C_{l_2}^2, C_{l_3}^2, ..., C_{l_n}^2, C_{l_1}^2\}$. ES in the remaining rounds are determined in the same way. Then the two subsequences interleave with each other. All channels in the odd time slots of CH sequence follow Odd Subsequence (OS) while all channels in the even time slots of CH sequence follow Even Subsequence (ES). Fig. 3(c) shows the interleaved final CH sequence of receiver-role. The last column means the channel in time slot $T$. If $T$ is odd, then the channel should be $OS_{T/2+1}$. If $T$ is even, then the channel should be $ES_{T/2}$. Fig. 4(a), (b) show the OS and ES when $C^2=\{1, 3, 4\}$ and $\{C_{l_1}^2, C_{l_2}^2, ..., C_{l_n}^2\}=\{3, 4, 1\}$. Fig. 4(c) shows the interleaved CH sequence. The channels with underline are from OS. Others are from ES.
\begin{figure*}
\centering
\subfigure[Structure of Odd Subsequence (OS)]{
\label{fig:subfig:a} %% label for first subfigure
\includegraphics[scale=1.6]{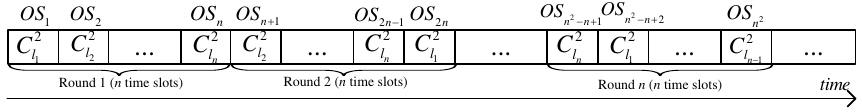}}
\hspace{1in}
\subfigure[Structure of Even Subsequence (ES)]{
\label{fig:subfig:b} %% label for second subfigure
\includegraphics[scale=1.6]{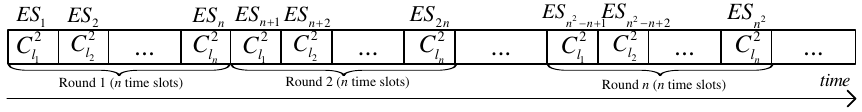}}
\subfigure[Interleave OS and ES]{
\label{fig:subfig:c} %% label for second subfigure
\includegraphics[scale=2]{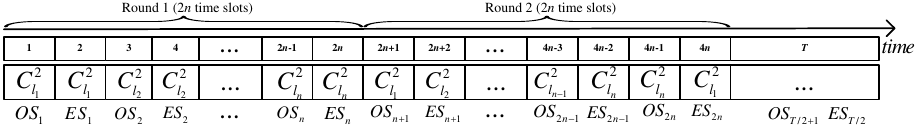}}
\caption{Structure of Receiver-role Sequence.}
\label{fig:subfig} %% label for entire figure
%\vspace{-0.2in}
\end{figure*}
\begin{figure}
\centering
\subfigure[Odd Subsequence]{
\label{fig:subfig:a} %% label for first subfigure
\includegraphics[scale=0.65]{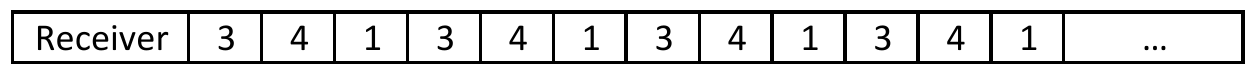}}
\hspace{1in}
\subfigure[Even Subsequence]{
\label{fig:subfig:b} %% label for second subfigure
\includegraphics[scale=0.65]{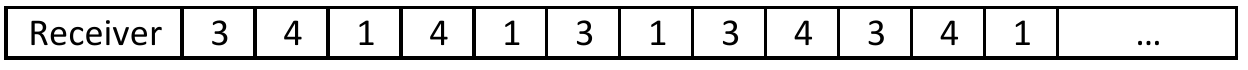}}
\subfigure[Interleave OS and ES]{
\label{fig:subfig:c} %% label for second subfigure
\includegraphics[scale=0.65]{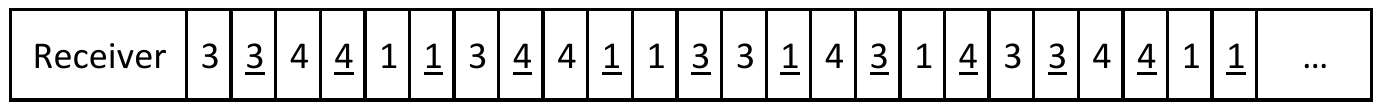}}
\caption{Receiver-role Sequence when $C^2=\{1, 3, 4\}$.}
\label{fig:subfig} %% label for entire figure
%\vspace{-0.2in}
\end{figure}
\par The CH sequence of sender and the OS of receiver can guarantee the rendezvous when $n=km_p$ ($k\geq1$). The CH sequence of sender and the ES of receiver can guarantee the rendezvous when $n\neq km_p$ ($k\geq1$).The ISAC algorithm is formally presented as Algorithm 1 and Algorithm 2.
\begin{algorithm}
\caption{ISAC algorithm (for generating sender-role sequence)}
\label{alg:Random Algorithm}
%\algsetup{indent=1em}
\begin{algorithmic}[1]
\REQUIRE sender's available channel $C^1=\{C_1^1, C_2^1, ..., C_m^1\}$
\STATE $m_p=$ the smallest prime number not smaller than $m$
\STATE Randomly select $m_p-m$ channels from $C^1$ and denote these channels by $\{C_{m+1}^1, C_{m+2}^1, ..., C_{m_p}^1\}$
\STATE Randomly select an integer $k$ in $[1, m_p]$
\STATE $t=0$
\WHILE{not rendezvous}
\STATE $t=t+1$
\STATE $T=(t-2+k)\%m_p +1$
\STATE Attempt rendezvous on channel $C_T^1$
\ENDWHILE
\end{algorithmic}
\end{algorithm}
\begin{algorithm}
\caption{ISAC algorithm (for generating receiver-role sequence)}
\label{alg:Random Algorithm}
%\algsetup{indent=1em}
\begin{algorithmic}[1]
\REQUIRE receiver's available channel $C^2=\{C_1^2, C_2^2, ..., C_n^2\}$
\STATE Randomly generate a permutation of $C^2: \{C_{l_1}^2, C_{l_2}^2, ..., C_{l_n}^2\}$
\STATE $t=0$
\WHILE{not rendezvous}
\STATE $t=t+1$
\IF{$t\%2==1$}
\STATE $T=(\frac{t}{2})\%n +1$\quad\quad\quad\quad\quad\quad\quad\quad\quad\quad//OS
\ELSE
\STATE $T=((\frac{t-1}{2n})\%n +(\frac{t}{2})\%n -1)\% n+1$\quad\quad//ES
\ENDIF
\STATE Attempt rendezvous on channel $C_{l_T}^2$
\ENDWHILE
\end{algorithmic}
\end{algorithm}
\par In the algorithm for sender, line 3 randomly selects a starting channel. Line 7 calculates the corresponding channel in time slot $t$ in round-robin fashion. In the algorithm for receiver, line 5 indicates whether the current time slot is odd time slot or not. Line 6 calculates the corresponding channel in time slot $t$ if $t$ is an odd. Line 8 calculates the corresponding channel in time slot $t$ when $t$ is an even.
\par \emph{An illustration example:} Available channel sets of sender and receiver are $C^1=\{1, 2\}$ and $C^2=\{1, 3, 4\}$, respectively. That is, $m_p=m=2$ and $n=3$. Fig. 5 shows the rendezvous of the two users. Notice that time-synchronization is not available. Since the sender has only two available channels, the \emph{offset} between the two users' sequences could be 0 or 1 (as shown in Fig.5). Sender and receiver can achieve rendezvous regardless of the offset between their sequences (see the time slots in gray color in Fig. 5).
\begin{figure}
\centering
\subfigure[Offset=1]{
\label{fig:subfig:a} %% label for first subfigure
\includegraphics[scale=0.63]{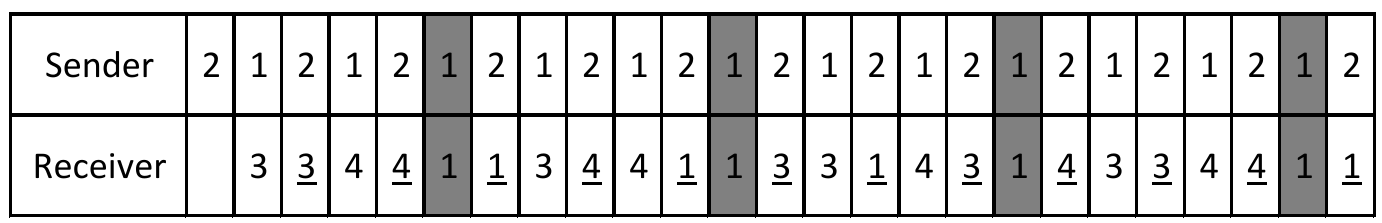}}
\hspace{1in}
\subfigure[Offset=0]{
\label{fig:subfig:b} %% label for second subfigure
\includegraphics[scale=0.63]{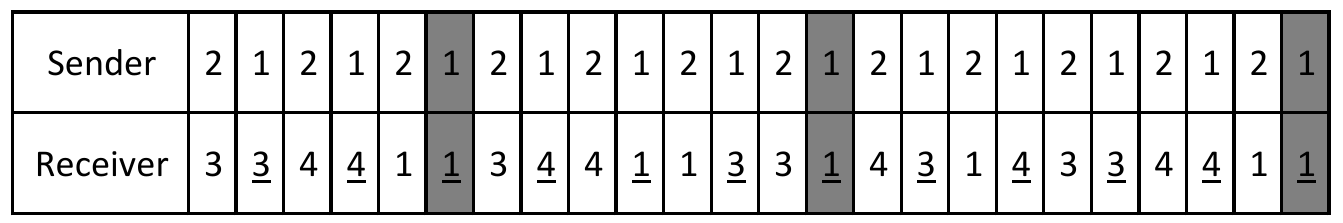}}
\caption{Rendezvous of two users performing ISAC when $C^1=\{1, 2\}$ and $C^2=\{1, 3, 4\}$.}
%\vspace{-0.2in}
\end{figure}
\subsection{Algorithm Analysis}
\par The ISAC algorithm has an important feature that distinguishes itself from the existing algorithms using the available channel set: it guarantees that rendezvous can be completed within finite time (see Table I). In this section, we prove this feature and derive the upper bounds of MTTR under the symmetric model and the asymmetric model. We first present the following two lemmas.
\begin{lemman}
Given two positive integers $x$ and $y$, and $y$ is a prime number. The statement ``$x$ is not evenly divisible by $y$" is equivalent to that ``$x$ and $y$ are co-prime".
\end{lemman}
\begin{proof}
$\Longrightarrow$Suppose $z$ is a common factor of $x$ and $y$. Since $y$ is a prime number, $z$ can be only 1 or $y$ itself. However, if $z$ is identical to $y$, then $x$ must be evenly divisible by $y$, leading to a contradiction. Thus, $x$ and $y$ have no other common factors rather than 1, i.e., they are co-prime.
\par$\Longleftarrow$Suppose that $x$ is evenly divisible by $y$, i.e., there exists $z$ such that $x=zy$. Since $y$ is a prime number, $y$ is not equal to 1. Thus, 1 and $y$ are both common factors of $x$ and $y$, i.e., $x$ and $y$ are not co-prime, leading to a contradiction. So, $x$ cannot be evenly divided by $y$.                                                               \end{proof}
\begin{lemman}
In the sender-role sequence that is generate by Algorithm 1, any consecutive $m_p$ channel indices in odd time slots (e.g., channel indices in time slots $1, 3, 5, ...$) form a permutation of the channel indices in $C^1_*=\{C_1^1, C_2^1, ..., C_m^1, C_{m+1}^1, C_{m+2}^1, ..., C_{m_p}^1\}$. This statement also holds for the channel indices in even time slots.
\end{lemman}
\begin{proof}
Suppose the user has $m$ available channels and the $m$ channels $C_m^1$ has been expanded to $C_{m_p}^1$. The starting channel index is $k$. Then the channel index in the time slot $T$ can be denoted as $(T-1+k)\%m_p$. Any  consecutive $m_p$ channel indices in odd time slots are $O_1\%m_p, (O_1+2)\%m_p, ..., (O_1+2m_p)\%m_p$ if the first channel of these channels is $C_{O_1}$. To prove that this subsequence is a permutation of $m_p$ channel index, we refer to Lemma 1 in [5], ``Given a positive integer $P$, if integer $r\in[1, P)$ is relatively prime to $P$ (i.e., the common factor between them is 1), then for any integer $x\in[0, P)$ the sequence $S=<(x~mod~P)+1, ((x+r)~mod~P)+1, ..., ((x+(P-1)r)~mod~P)+1>$ is a permutation of $<1, 2, ..., P>$". In our analysis, $r=2$ is relatively prime to $P=m_p$. So $O_1+0\%m_p, (O_1+2)\%m_p, ..., (O_1+2(m_p-1))\%m_p$ is a permutation of $<1, 2, ..., m_p>$.
\end{proof}
\par Based on Lemma IV.1 and Lemma IV.2, we prove the correctness of ISAC and derive its upper bounds of MTTR under both the symmetric model and the asymmetric model.
\subsubsection{Under the Symmetric Model}
Under the symmetric model, the sender and the receiver have the same available channels. $G=m=n$.
\newtheorem{theorem}{Theorem}
\begin{theoremn}
 Under the symmetric model, two users performing the ISAC algorithm can achieve rendezvous in at most $2m_p-1$ time slots, where $m$ and $n$ are the numbers of channels of the two users, $m_p$ is the smallest prime number which is not smaller than $m$.
\end{theoremn}
\begin{proof}
\begin{figure*}
\centering
\subfigure[Sender-role Sequence and Receiver-role Sequence]{
\label{fig:subfig:a} %% label for first subfigure
\includegraphics[scale=2]{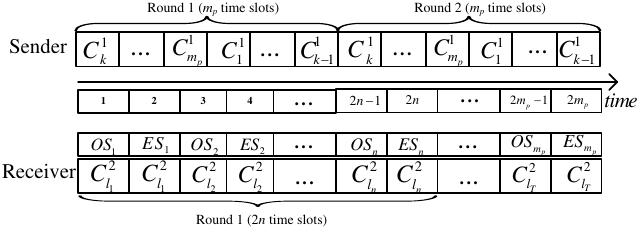}}
\hspace{1in}
\subfigure[Rendezvous achieved between Sender-role sequence and ES of Receiver-role sequence]{
\label{fig:subfig:b} %% label for second subfigure
\includegraphics[scale=2]{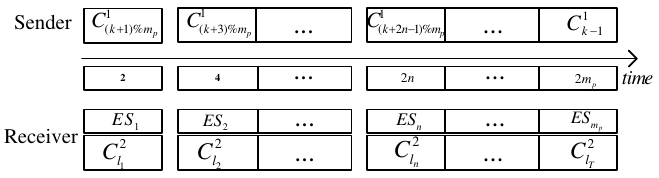}}
\caption{Rendezvous of two users performing ISAC.}
%\vspace{-0.2in}
\end{figure*}
We consider the CH sequence of sender and the Odd Subsequence of receiver. Then we investigate the consecutive time-span of $2m_p$ time slots from the starting point. Fig. 6(a) shows the CH sequence of sender and receiver. Fig. 6(b) extracts the channels in the odd time slots from Fig. 6(a). We now prove that there is a guaranteed rendezvous between the sequence of sender and OS of receiver. Suppose that the available channels of two users are $\{C_1, C_2, ..., C_n\}$. The first channel on the extracted sequence of sender is $C_i$ and the first channel on the OS of receiver is $C_j$. According to the ISAC algorithm for sender and receiver, the $k\text{-}$th channel on the extracted sequence of sender is $C_{i+2k\%m_p}$ while the OS of receiver is $C_{(j+k)\%n}$. For all $1\leq i\leq m_p$, $1\leq j\leq n$, we must find a $k$ ($0\leq k\leq m_p-1$) such that $(i+2k)\%m_p=(j+k)\%n$. We should have
\begin{eqnarray}
%\begin{flalign}
&&i+2k=a+bm_p\\
&&j+k=a+cn
%\end{flalign}
\end{eqnarray}
where $a\in[0, n)$, $b\in\{0, 1, 2\}$, and $c\in\{0, 1\}$. Then we have
\begin{eqnarray}
&&k=bm_p-cn+j-i
\end{eqnarray}
Since there is no time-synchronization, $(j-i)\in(-m_p,+m_p)$. We divide it into three subcases.
\begin{itemize}
\item \emph{Subcase 1:} $(j-i)\in(-m_p, 0)$, we let $b=1$ and $c=0$. $k=m_p+j-i$ and $0<k<m_p$;
\item \emph{Subcase 2:} $(j-i)\in[0, n)$, we let $b=1$, and $c=1$. $k=m_p-n+j-i$ and $m_p-n\leq k<m_p$;
\item \emph{Subcase 3:} $(j-i)\in[n, m_p)$, we let $b=0$, and $c=1$. $k=-n+j-i$ and $0\leq k<m_p-n$;
\end{itemize}
Therefore, we can always find at least one pair $(b, c)$ such that $0\leq k\leq m_p-1$. There is a guaranteed rendezvous before $m_p$ time slots in the Odd Subsequence which corresponds to $2m_p-1$ time slots in the whole CH sequence.
\end{proof}
\subsubsection{Under the Asymmetric Model}
Under the asymmetric model, sender and receiver may have different available channels. Since the case of $G=m=n$ has already been proved, here we suppose that $G\neq m$ or $G\neq n$.
\begin{theoremn}
Under the asymmetric model, two users performing the ISAC algorithm can achieve rendezvous in at most $2m_pn-2G+2$ time slots, where $m$ and $n$ are the numbers of channels of the two users, $m_p$ is the smallest prime number which is not smaller than $m$, and $G$ is the number of commonly-available channels of the two users.
\end{theoremn}
\begin{proof}
We prove the correctness of ISAC under the asymmetric model in two cases: 1) $n$ is not divisible by $m_p$; 2) $n$ is divisible by $m_p$ but $G\neq m$ or $G\neq n$.
\par If $n$ is not divisible by $m_p$ and $m>2$, we consider the CH sequence of sender and the Odd Subsequence of receiver. Based on Fig. 3 which shows the structure of sequences of ISAC and Lemma IV.2, we extract the channels in the odd time slots of sender-role sequence and denote this subsequence in round 1 by $S^1=\{C_k^1,C_{k+1}^1, ...,C_{m_p}^1, C_1^1, C_2^1, ..., C_{k-1}^1\}$ (with $m_p$ items). The receiver's OS in round 1 is $S^2=\{C_{l_1}^2, C_{l_2}^2, ..., C_{l_n}^2\}$ (with $n$ items). Furthermore, we use $S_h^1$ ($S_h^2$) to denote the $h\text{-}$th item of $S^1$ ($S^2$) ($h=1, 2, ...$). Then, according to ISAC, in time slot $(2T-1)$ sender and receiver actually hop on channel $S_{((T-1)\%m_p)+1}^1$ and channel $S_{((T-1)\%n)+1}^2$, respectively. Since there is no time-synchronization, without loss of generality, we investigate rendezvous of the two users at a starting point in which sender and receiver are in time slots $t_0^1$ and $t_0^2$, respectively. Next, we prove that, for any pair of channels $S_h^1\in S^1$ and $S_g^2\in S^2$, there must be a same time slot at which sender and receiver hop on $S_h^1$ and $S_g^2$, respectively. Notice that this result implies the guaranteed rendezvous of the two users since they have commonly-available channels (i.e., $S^1$ and $S^2$ share common items). To achieve this, we should prove that, for any $1\leq h\leq m_p$ and $1\leq g\leq n$, there exists $t$ to satisfy the following equations (i.e., sender hops on channel $S_h^1$ in its time slot $t+t_0^1$ and receiver hops on channel $S_g^2$ in its time slot $t+t_0^j$)
\begin{eqnarray}
&&h-1\equiv t+t_0^1-1~mod~(m_p)\\
&&g-1\equiv t+t_0^2-1~mod~(n)
\end{eqnarray}
which can be equivalently rewritten as follows
\begin{eqnarray}
%\begin{aligned}
&&t\equiv (h-t_0^1)~mod~(m_p)\\
&&t\equiv (g-t_0^2)~mod~(n)
%\end{aligned}
\end{eqnarray}
\par Since $m_p$ is a prime number and $n$ is not evenly divisible by $m_p$, from Lemma IV.1, we know $m_p$ and $n$ are co-prime. Therefore, from the Chinese Remainder Theorem [9], there exists an integer $t$ that solves Equations (6) and (7). As mentioned earlier, this result implies the guaranteed rendezvous of the two users. We further derive an upper bound of TTR as follows. Notice that there are $m_pn$ combinations of $h$ and $g$ values (i.e., $m_pn$ channel pairs of $S_h^1$ and $S_g^2$). Hence, $t$ does not exceed $m_pn$ regardless of $h$ and $g$ values. On the other hand, since the two users have $G$ commonly-available channels (i.e., $S^1$ and $S^2$ share $G$ common items), there are $G$ pairs of commonly-available channels among all the $m_pn$ pairs of $S_h^1$ and $S_g^2$. Therefore, rendezvous will occur in at most $m_pn-G+1$ time slots in the Odd Subsequence and the corresponded time slot in the whole CH sequence is $2m_pn-2G+1$. In other words, $TTR\leq 2m_pn-2G+1$.
\par We consider the next case that $n$ is divisible by $m_p$ but $G\neq m$ and $G\neq n$.
\par If $n$ is divisible by $m_p$, we consider CH of sender and the Even Subsequence of receiver. Then, we investigate any consecutive time-span of $n\times m_p$ time slots from a starting point in which sender and receiver are in time slots $t_0^1$ and $t_0^2$.
\par We prove that, any pair of channels $C_h^1$ in $\{C_1^1, C_2^1, ...,C_m^1, C_{m+1}^1, ..., C_{m_p}^1\}$ and $C_g^2$ in $\{C_{l_1}^2, C_{l_2}^2, ..., C_{l_n}^2\}$ appears once and only once in the time-span. That is, there exists exactly one $t$ ($0<t\leq n\times m_p$) such that sender hops on channel $C_h^1$ in its time slot $t_0^1+t$ and receiver hops on channel $C_g^2$ in its time slot $t_0^2+t$. Next we prove it by contradiction.
\par Suppose that a pair of channels, say $C_h^1$ and $C_g^2$ appears more than one time. This implies that there exist two different $t_1$ and $t_2$ such that:
\begin{center}
sender hops on channel $C_h^1$ in time slot $t_0^1+t_1$,\\
receiver hops on channel $C_g^2$ in time slot $t_0^2+t_1$,
\end{center}
and
\begin{center}
sender hops on channel $C_h^1$ in time slot $t_0^1+t_2$,\\
receiver hops on channel $C_g^2$ in time slot $t_0^2+t_2$,
\end{center}
\par According to our ISAC algorithm and Lemma IV.2, we must have (suppose that $t_2>t_1$ for convenience):
\begin{eqnarray}
t_2=t_1+\mu\times m_p
\end{eqnarray}
\par Sender-role hops on the $m_p$ channels in round-robin fashion. The channel will appear after each $m_p$ time slots. Each channel of sender will appears for $n$ times in any continuous $n\times m_p$ time slots. There are at most another $n-1$ rounds after the first time this channel appears, so $1\leq\mu<n$.
\begin{eqnarray}
t_2=t_1+a\times(n-1)+b\times n
\end{eqnarray}
\par In each round of receiver's ES, all channel indices are left-shifted by 1 from the previous round. The channel will appear after each $(n-1)$ time slots. However, when the channel is the first one in current round, it will appear after $(n-1)+n$ time slots (the third and fourth channel 1 in Fig. 4), so $b\in\{0,1\}$.  Each channel of sender will appears for $m_p$ times in any consecutive $n\times m_p$ time slots. There are at most another $n-1$ rounds after the first time this channel appears, so $1\leq a<m_p$.
\par Using Equations (8) and (9), and the fact that $n=\delta\times m_p$ ($n$ is evenly divisible by $m_p$), we should have
\begin{eqnarray}
&&\mu\times m_p=a\times (\delta\times m_p-1)+b\times\delta\times m_p\\
&&a=((a+b)\times\delta-\mu)\times m_p
\end{eqnarray}
\par That is, $a$ should be divisible by $m_p$, which contradicts the requirement $1\leq a<m_p$. So any pair of channels $C_h^1$ and $C_g^2$ in any consecutive time-span of $n\times m_p$ time slots appears once and only once.
\par This result implies the guaranteed rendezvous of the two users. Notice that there are $m_p\times n$ combinations of $h$ and $g$ values (i.e., $m_pn$ channel pairs of $S_h^1$ and $S_g^2$) in at most $m_pn$ time slots. Hence, the combinations in $m_pn$ time slots are different to each other. We further derive an upper bound of TTR as follows. If there are $G$ commonly-available channels of the two users, there are $G$ pairs of commonly-available channels among all the $m_pn$ pairs of $S_h^1$ and $S_g^2$. The worst case is that the whole $G$ times rendezvous occur in the last $G$ time slots of channel hopping sequence. Therefore, rendezvous will occur in at most $m_pn-G+1$ time slots in the Even Subsequence and the corresponded time slot in the whole CH sequence is $2m_pn-2G+2$. In other words, $TTR\leq 2m_pn-2G+2$.
\par Then we illustrate the case when $m=1$ and $m=2$. If $m=1$, the only channel of sender must be a commonly-available channel of the two users. The sender stay on the only channel while the receiver will hop on this channel in at most $2n-1$ time slots. If $m=2$, all channels in the odd time slots of sender are one channel and all channels in the even time slots are another channel. No matter which one is a commonly-available channel, the receiver will hop on this channel in at most $n-G+1$ time slots in its OS or ES while the corresponded time slot in the whole CH sequence is $2n-2G+2$.
\par To sum up, two users performing the ISAC algorithm can achieve rendezvous in at most $(2m_pn-2G+2)$ times lots, where $m$ and $n$ are the numbers of channels of the two users, $m_p$ is the smallest prime number which is not smaller than $m$, $G$ is the number of commonly-available channels of the two users.
\end{proof}
\section{Simulation}
We built a simulator in Visual Studio 2010 to evaluate the performance of our proposed ISAC algorithm. We consider the following algorithms for comparison: i) MMC [9] (MMC generates CH sequences based on available channel set), ii) Jump-Stay [5] [6] and iii) CRSEQ [8] (Among existing algorithms, Jump-Stay and CRSEQ have been shown to have good performance [7]). We remind that Jump-Stay and CRSEQ apply a random replacement operation to randomly replace the unavailable channels in the CH sequences by the available ones (see discussion in Section I). We introduce a parameter $\theta~(0<\theta\leq1)$ to control the ratio of the number of available channels to the total number of channels $Q$. Available channels are randomly selected from the whole channel set such that the average number of available channels is equal to $\theta Q$. Besides $\theta$, we are concerned with parameter $G$, i.e., the number of commonly-available channels of the two users involved in the rendezvous. For each $\theta$, we let $G$ properly vary in $[1,\theta Q]$. For each combination of parameter values, we perform 500,000 independent runs and compute the average TTR, the maximum TTR (MTTR) and the variance of TTR accordingly. The variance of TTR can reveal the stability of performance of rendezvous algorithms. Notice that, $n$ is the number of available channels of receiver, and $m_p$ is the smallest prime number which is not smaller than the number of available channels of sender.
\subsection{Effectiveness of ISAC}
In this subsection, we demonstrate that  ISAC can effectively improve the rendezvous performance. We consider three scenarios with different number of available channels: 1) $\theta$ is small and equal to 0.1 (i.e., ratio of the number of available channels to the total number of channels is 0.1), ii) $\theta$ is moderate and equal to 0.4, and iii) $\theta$ is large and equal to 0.8.
%\begin{itemize}
\subsubsection{Under the Symmetric Model}
\begin{itemize}
\item\emph{Small $\theta$}: Firstly we study the scenario when $\theta$ is small. We set $\theta=0.1$ and $G=0.1Q$. Fig. 7(a) shows the average TTRs and Fig. 7(b) shows the maximum TTRs of different algorithms against the total number of channels $Q$. Since the maximum TTR of MMC is much bigger than that of other algorithms, we draw one more graph to compare the maximum TTRs of other three algorithms (i.e., Fig. 7(c)). Fig. 7(d) shows the variance of different algorithms. There is no large gap between different algorithms in terms of average TTR. However, the ISAC has significant improvement on the maximum TTR and the variance of TTR. For example, we suppose that each time slot has duration of 20ms [5]. When there are 50 channels, the average TTRs of ISAC, MMC, Jump-Stay and CRSEQ are 4.20 slots (0.08s), 6.85 slots (0.14s), 5.00 slots (0.10s) and 4.99 slots (0.10s) while the maximum TTRs of ISAC, MMC, Jump-Stay and CRSEQ are 8 slots (0.16s), 393 slots (7.86s), 63 slots (1.26s) and 72 slots (1.44s), respectively. In this example, $m=n=0.1\times50=5$. According to Theorem IV.1, $MTTR\leq 2\times m_p-1=2\times5-1=9$. This result shows that the upper bound given by Theorem IV.1 is tight. The variances of TTR of ISAC, MMC, Jump-Stay and CRSEQ in this case are 6.57, 183.77, 20.03 and 24.02, respectively. The performance of ISAC is the most stable one.
\begin{figure}
\centering
\subfigure[Average TTR VS. $Q$]{
\label{fig:subfig:a} %% label for first subfigure
\includegraphics[scale=0.5]{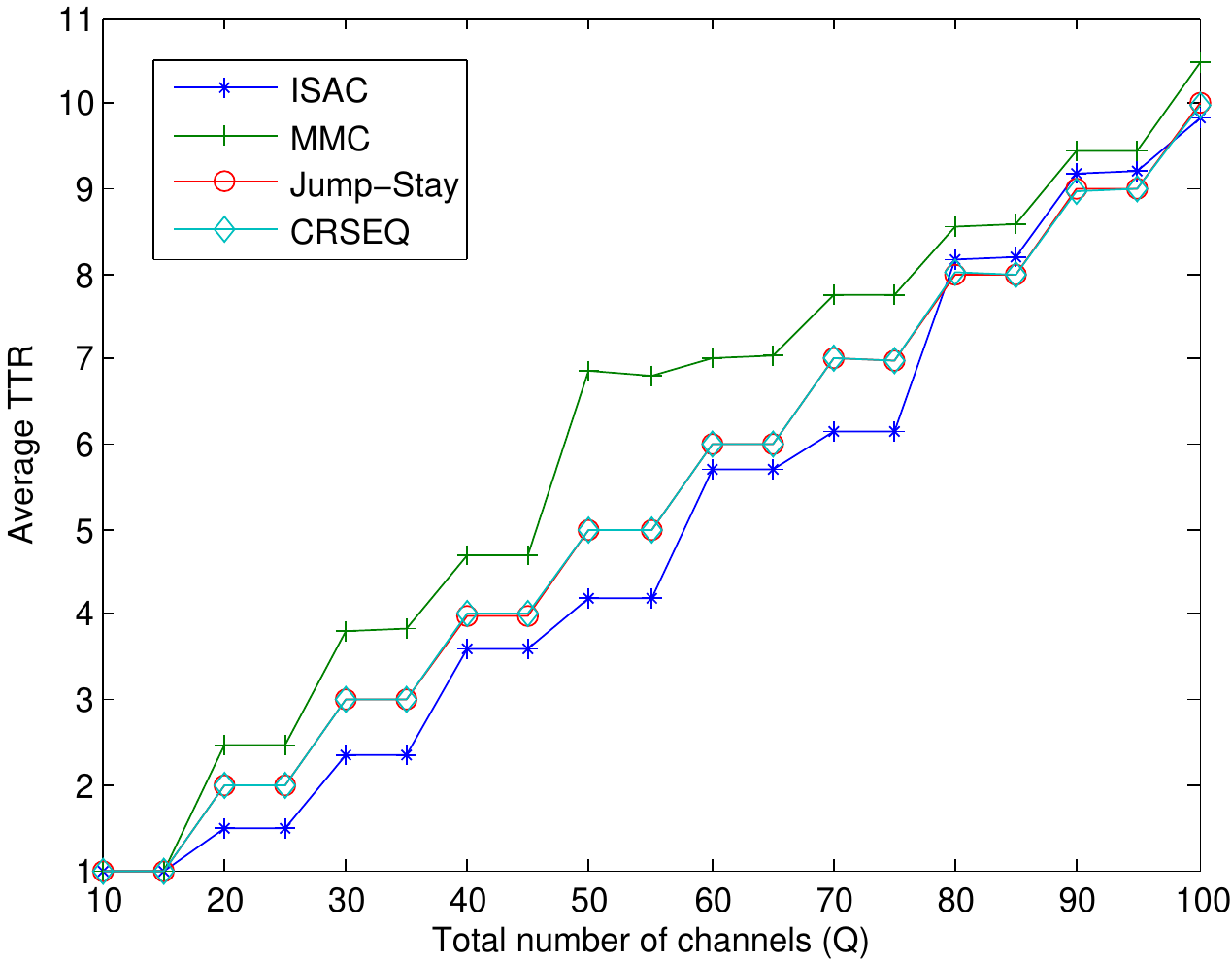}}
\hspace{1in}
\subfigure[Maximum TTR VS. $Q$]{
\label{fig:subfig:b} %% label for second subfigure
\includegraphics[scale=0.5]{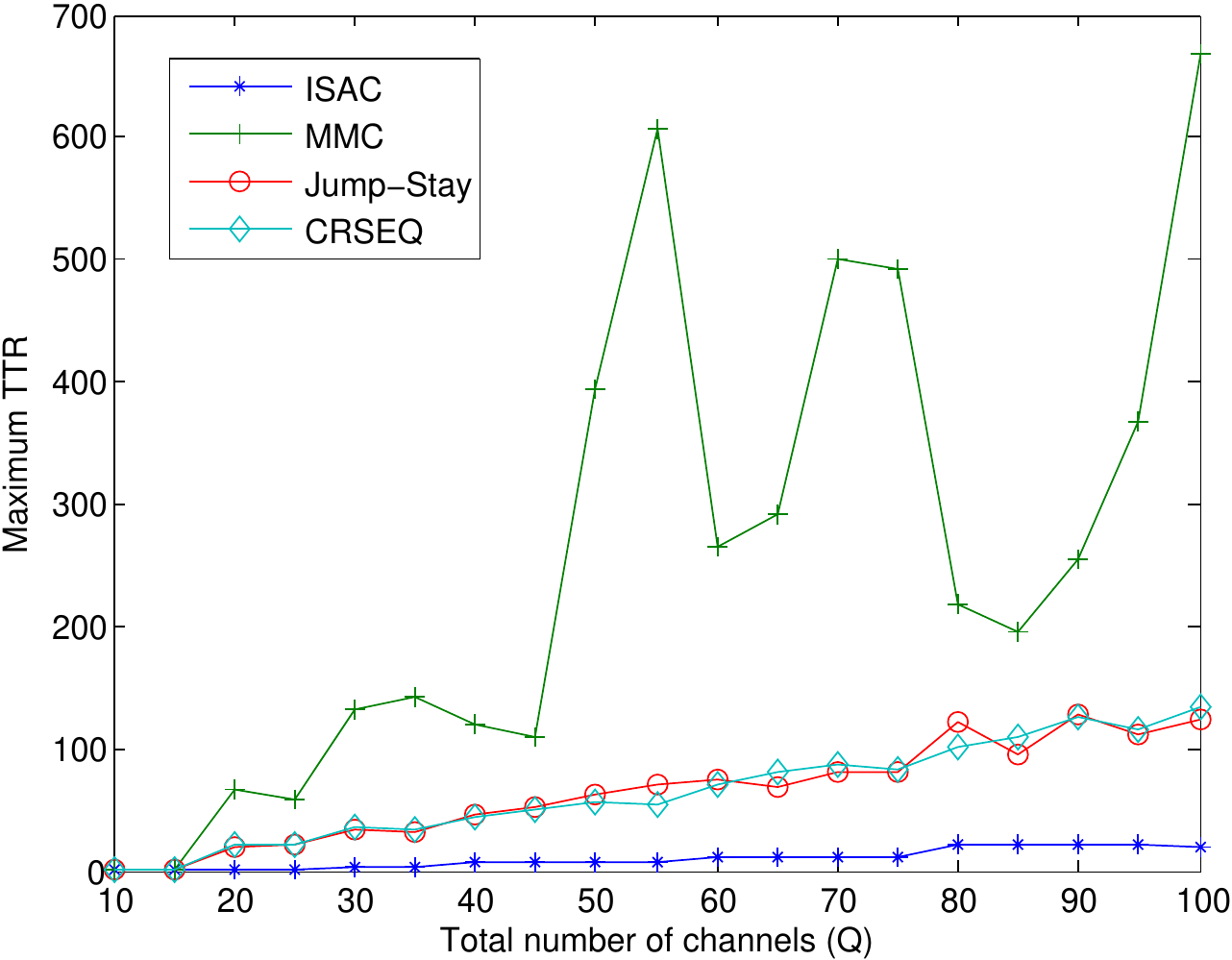}}
\hspace{1in}
\subfigure[Maximum TTR VS. $Q$ (without MMC)]{
\label{fig:subfig:b} %% label for second subfigure
\includegraphics[scale=0.5]{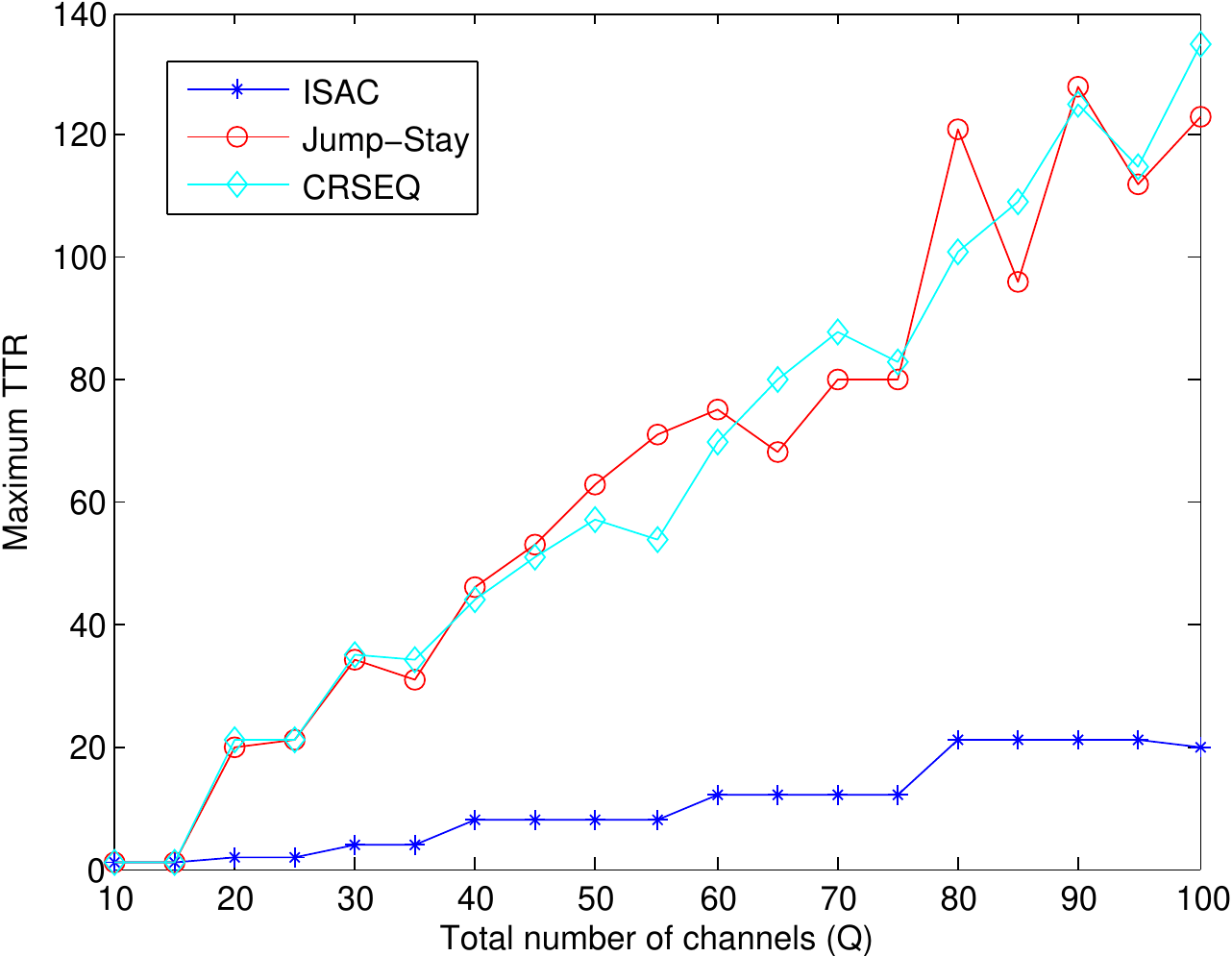}}
\hspace{1in}
\subfigure[Variance of TTR VS. $Q$]{
\label{fig:subfig:b} %% label for second subfigure
\includegraphics[scale=0.5]{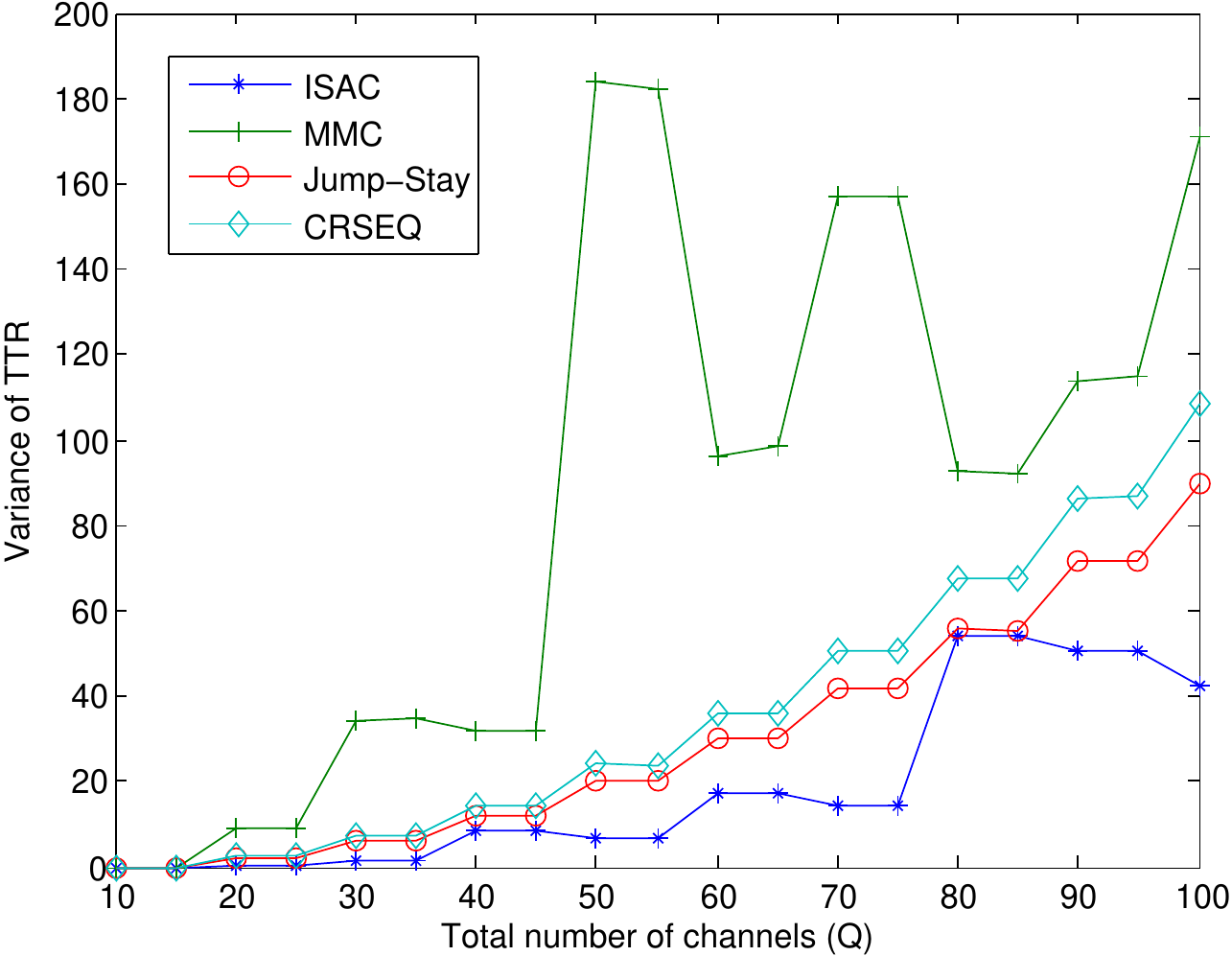}}
\caption{Comparison of different algorithms under the symmetric model when $\theta=0.1$.}
%\vspace{-0.2in}
\end{figure}
\item \emph{Moderate $\theta$}: Then we study the scenario when $\theta$ is moderate. We set $\theta=0.4$ and $G=0.4Q$. Fig. 8 shows the average TTRs, the maximum TTRs and the variances of TTR of different algorithms against $Q$. We find that ISAC has about the same average TTR as the existing algorithms, but its maximum TTR is much smaller than that of the existing algorithms. For example, when there are 50 channels, the maximum TTRs of ISAC, MMC, Jump-Stay and CRSEQ are 45, 870, 239 and 310, respectively. In this example, $m=n=0.4\times50=20$. According to Theorem IV.1, $MTTR\leq 2\times m_p-1=2\times 23\times-1=45$. Again, this result shows that the upper bound given by Theorem IV.1 is tight. The variances of TTR of ISAC, MMC, Jump-Stay and CRSEQ in this case are 208.99, 539.73, 351.34 and 526.13, respectively. The performance of ISAC is the most stable one.
\begin{figure}
\centering
\subfigure[Average TTR VS. $Q$]{
\label{fig:subfig:a} %% label for first subfigure
\includegraphics[scale=0.5]{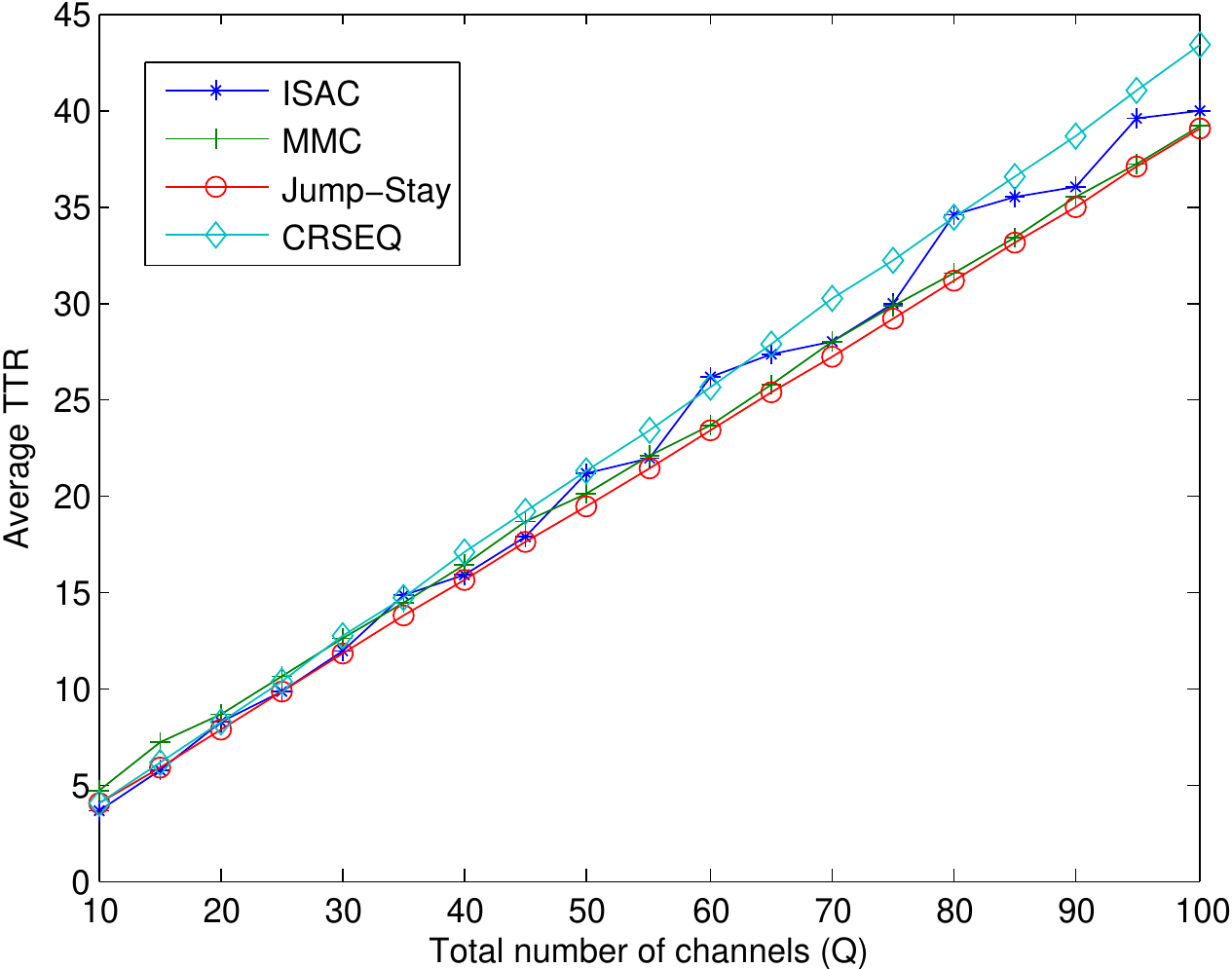}}
\hspace{1in}
\subfigure[Maximum TTR VS. $Q$]{
\label{fig:subfig:b} %% label for second subfigure
\includegraphics[scale=0.5]{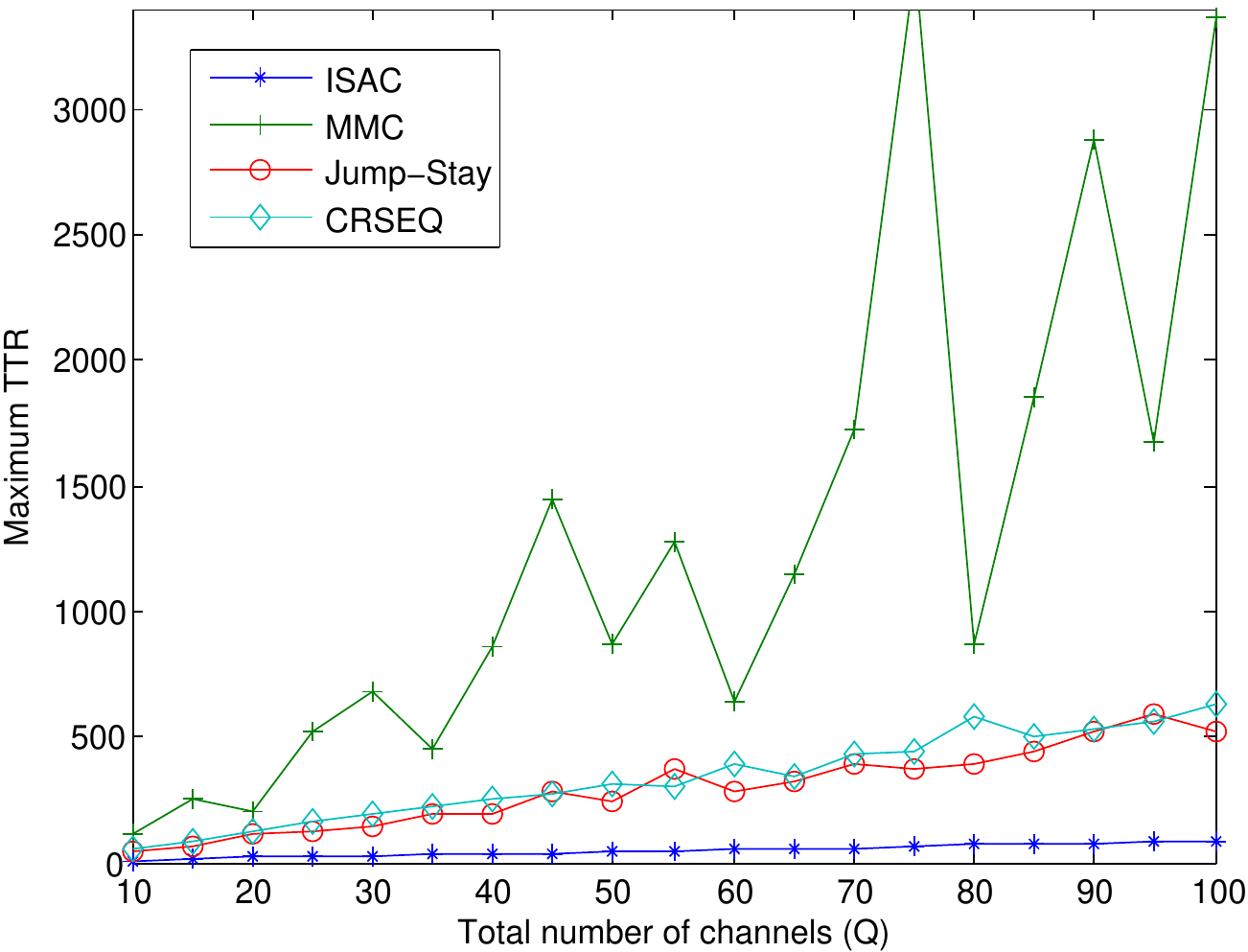}}
\hspace{1in}
\subfigure[Maximum TTR VS. $Q$ (without MMC)]{
\label{fig:subfig:b} %% label for second subfigure
\includegraphics[scale=0.5]{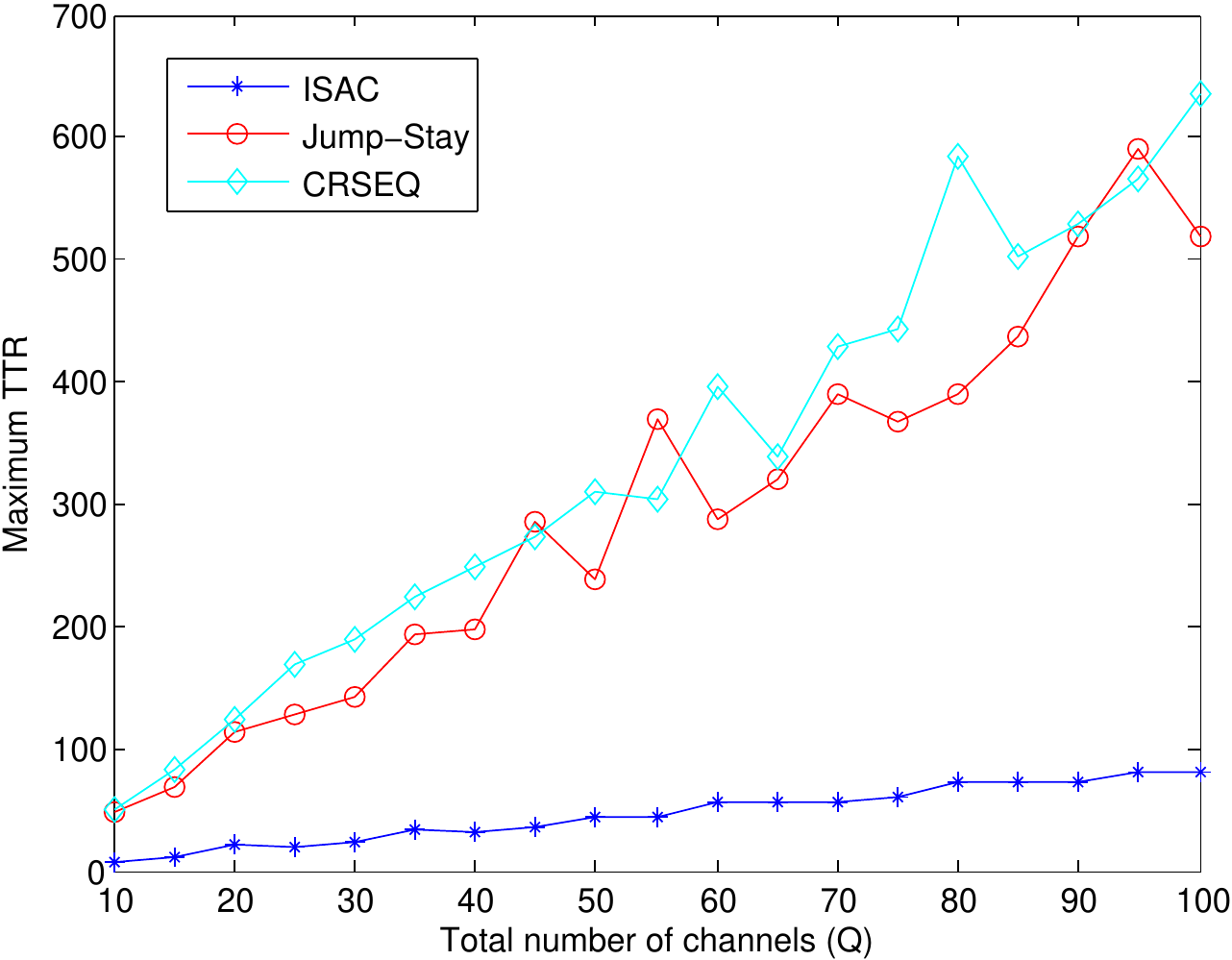}}
\hspace{1in}
\subfigure[Variance of TTR VS. $Q$]{
\label{fig:subfig:b} %% label for second subfigure
\includegraphics[scale=0.5]{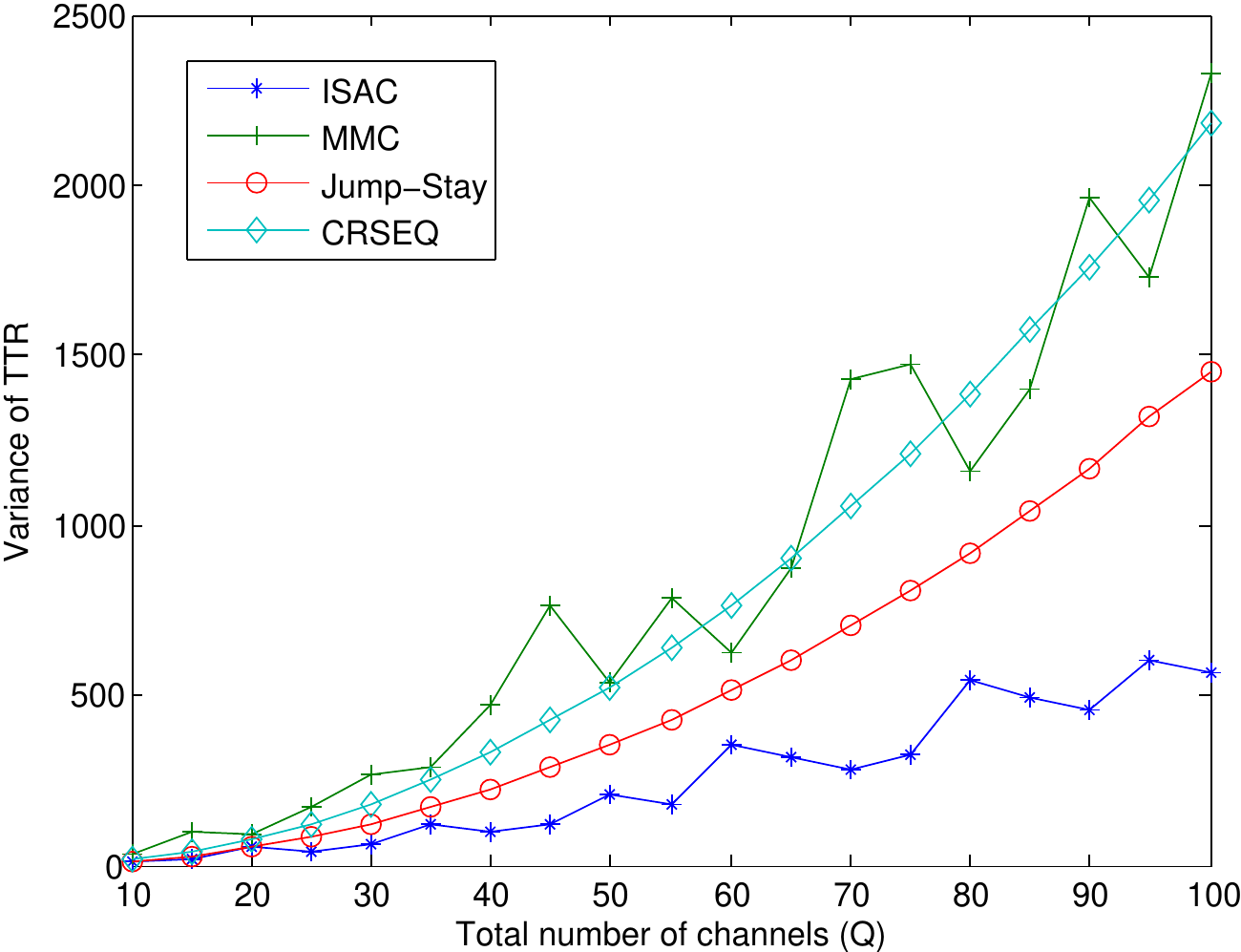}}
\caption{Comparison of different algorithms under the symmetric model when $\theta=0.4$.}
%\vspace{-0.2in}
\end{figure}
%\begin{figure}
%%\vspace{-0.25in}
%\mbox{
%\subfigure[Average TTR]{
%\begin{minipage}[b]{3.7cm}
%\includegraphics[width=1.7in, height=2in]{F10a.pdf}
%\end{minipage}
%}\quad
%\subfigure[Maximum TTR]{
%\begin{minipage}[b]{3.7cm}
%\includegraphics[width=1.7in, height=2in]{F10b.pdf}
%\end{minipage}}
%}
%\caption{Comparison of different algorithms under the symmetric model when $\theta=0.4$.}
%%\vspace{-0.15in}
%\end{figure}
\item \emph{Large $\theta$}: Then we study the scenario when $\theta$ is moderate. We set $\theta=0.8$, $G=0.8Q$. Fig. 9 shows the results. We see that ISAC is particularly advantageous in terms of the maximum TTR. For example, when there are 50 channels, the average TTRs of ISAC, MMC, Jump-Stay and CRSEQ are 39.92, 39.42, 33.29 and 58.65 while the maximum TTRs are 80, 3431, 489 and 1267, respectively. The maximum TTR of ISAC is almost one tenth of others. In this example, $m=n=0.8\times50=40$, $G=0.8\times50=40$. According to Theorem IV.1, $MTTR\leq 2\times m_p-1=2\times41-1=81$. Again, this result shows that the upper bound given by Theorem IV.1 is tight. The variances of TTR of ISAC, MMC, Jump-Stay and CRSEQ in this case are 564.05, 2495.31, 1098.35 and 4933.48, respectively. The performance of ISAC is the most stable one.
\begin{figure}
\centering
\subfigure[Average TTR VS. $Q$]{
\label{fig:subfig:a} %% label for first subfigure
\includegraphics[scale=0.5]{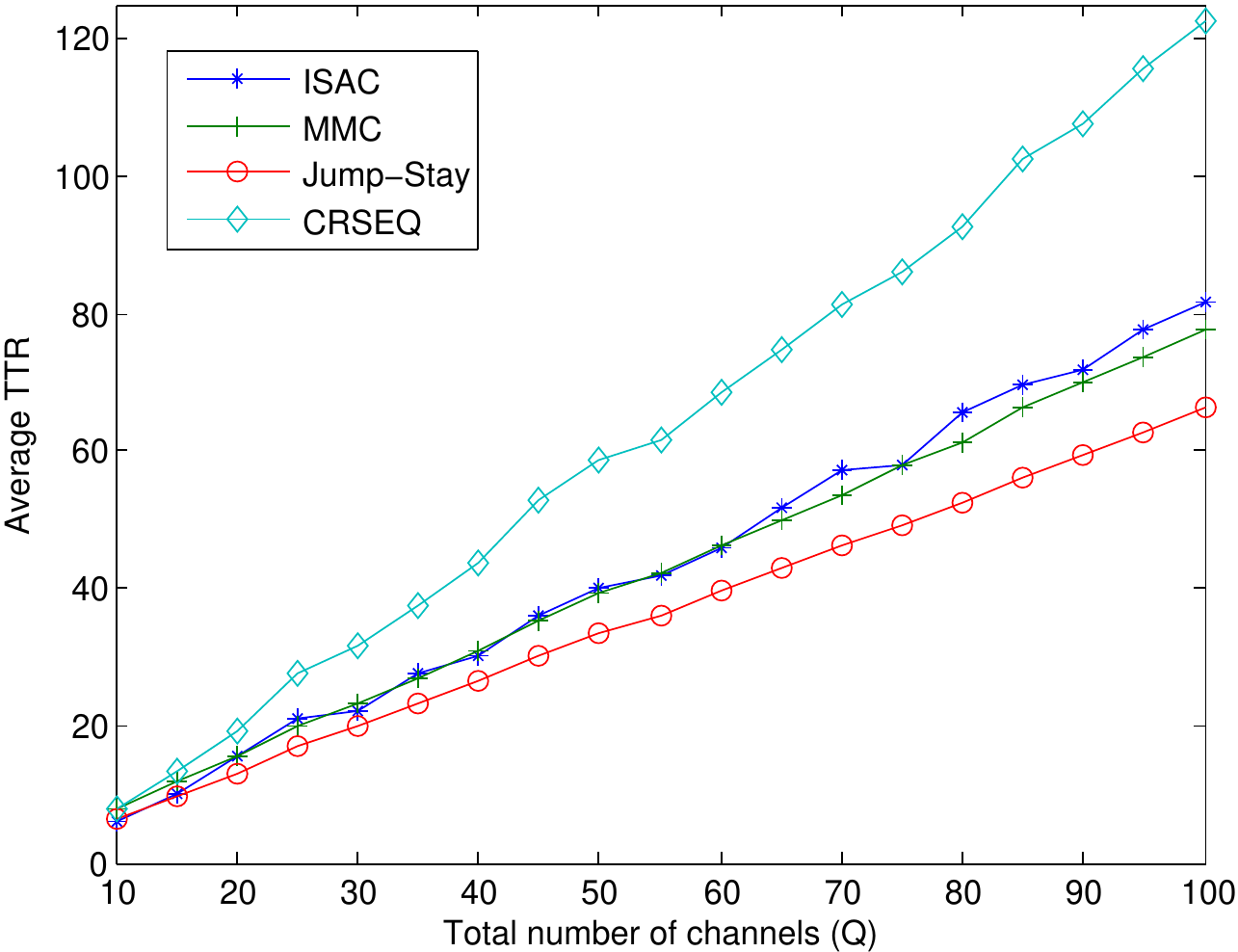}}
\hspace{1in}
\subfigure[Maximum TTR VS. $Q$]{
\label{fig:subfig:b} %% label for second subfigure
\includegraphics[scale=0.5]{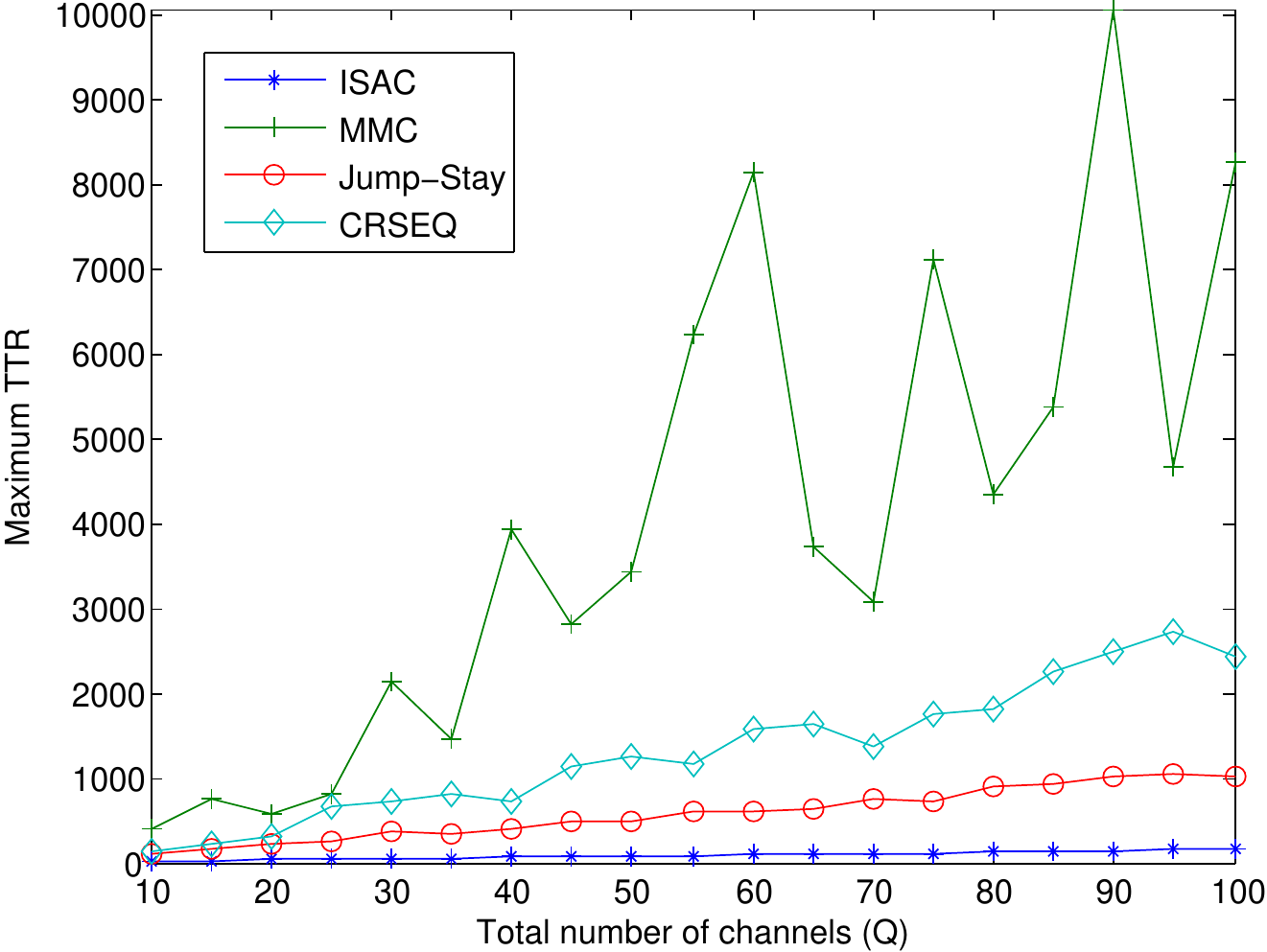}}
\hspace{1in}
\subfigure[Maximum TTR VS. $Q$ (without MMC)]{
\label{fig:subfig:b} %% label for second subfigure
\includegraphics[scale=0.5]{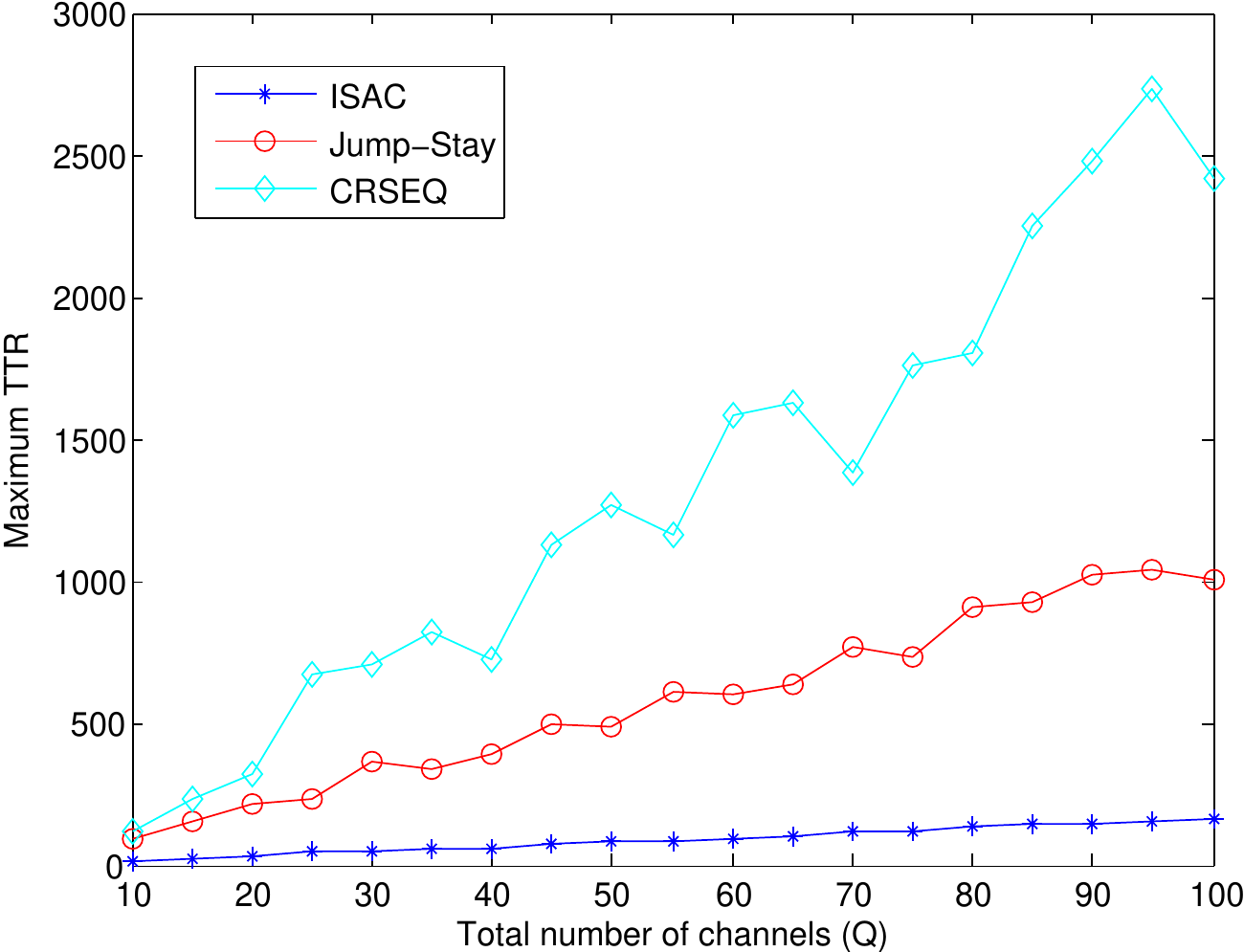}}
\hspace{1in}
\subfigure[Variance of TTR VS. $Q$]{
\label{fig:subfig:b} %% label for second subfigure
\includegraphics[scale=0.5]{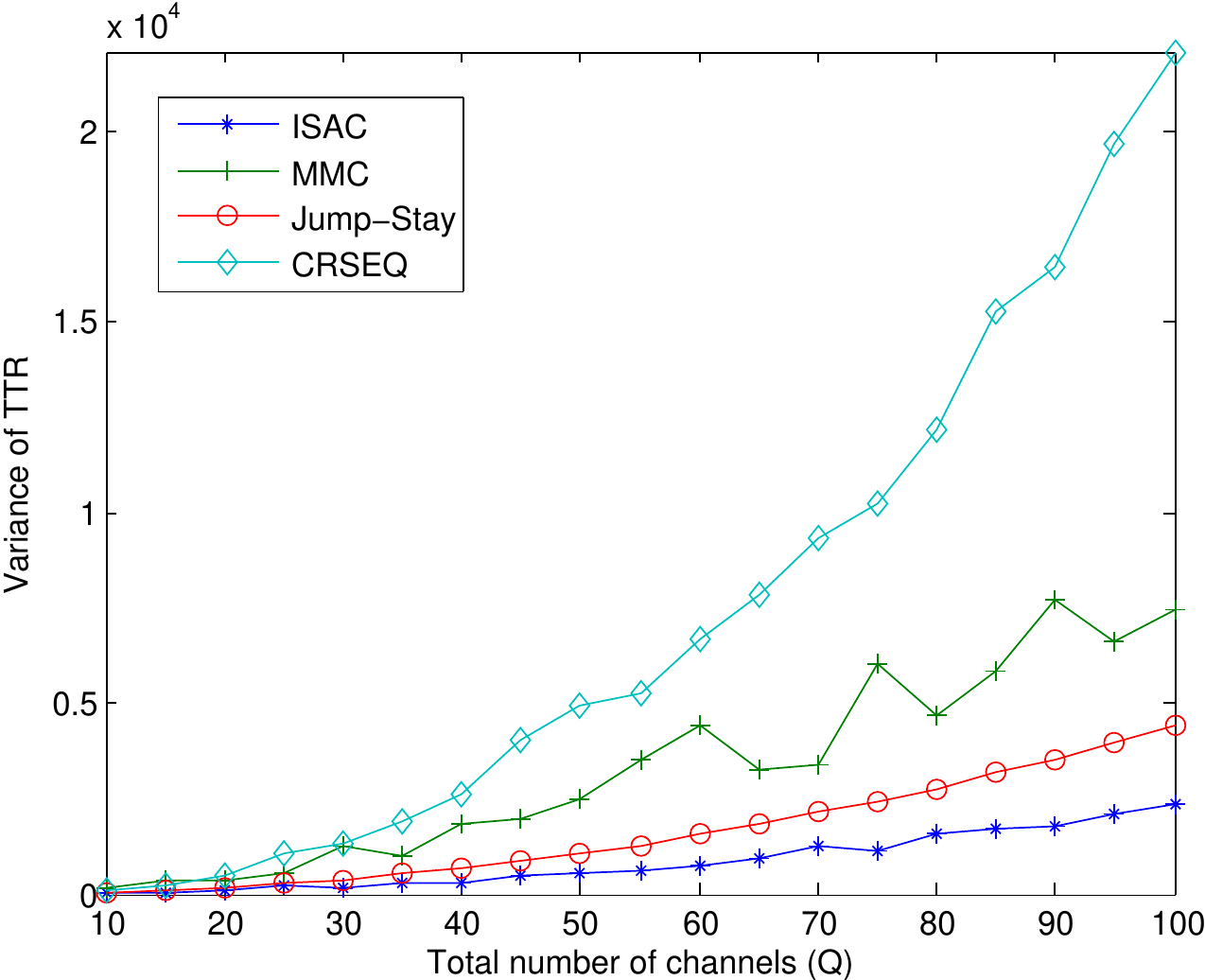}}
\caption{Comparison of different algorithms under the symmetric model when $\theta=0.8$.}
%\vspace{-0.2in}
\end{figure}
\end{itemize}
\subsubsection{Under the Asymmetric Model}
\begin{itemize}
\item \emph{Small $\theta$}: Firstly we study the scenario when $\theta$ is small, that is, only a small part of channels are available to users. We set $\theta=0.1$, $G=1$. Fig. 10 shows the average TTRs, the maximum TTRs and the variances of TTR of different algorithms against $Q$. Since there is only one commonly-available channel ($G=1$), such case is believed to be hard for quick rendezvous. According to Fig. 10, our ISAC algorithm significantly outperforms other algorithms in terms of both average TTR and maximum TTR. For example, when there are 50 channels, the average TTRs of ISAC, MMC, Jump-Stay and CRSEQ are 14.65, 27.00, 21.26 and 21.20 while the maximum TTRs are 98, 1023, 562 and 530, respectively. We find that ISAC has significant improvement on the maximum TTR. Its maximum TTR is almost one tenth of others. In this example, $m\leq0.15\times50=7.5$. According to Theorem IV.2, $MTTR\leq 2\times m_pn-2G+2=2\times 7\times7-2+2=98$. This result shows that the upper bound given by Theorem IV.2 is tight when $G=1$. The variances of TTR of ISAC, MMC, Jump-Stay and CRSEQ in this case are 255.06, 2080.64, 738.08 and 526.13, respectively. The performance of ISAC is the most stable one.
\begin{figure}
\centering
\subfigure[Average TTR VS. $Q$]{
\label{fig:subfig:a} %% label for first subfigure
\includegraphics[scale=0.5]{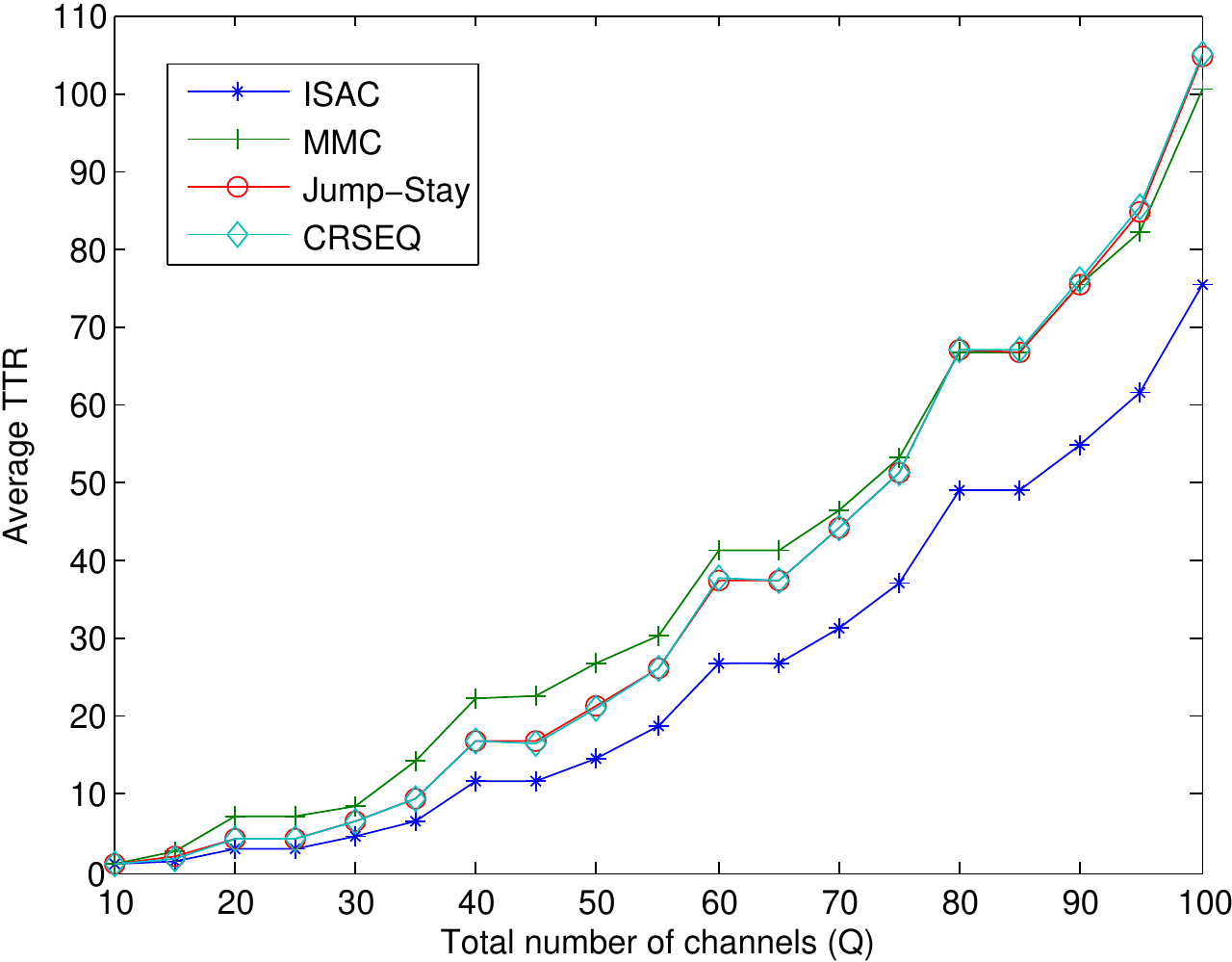}}
\hspace{1in}
\subfigure[Maximum TTR VS. $Q$]{
\label{fig:subfig:b} %% label for second subfigure
\includegraphics[scale=0.5]{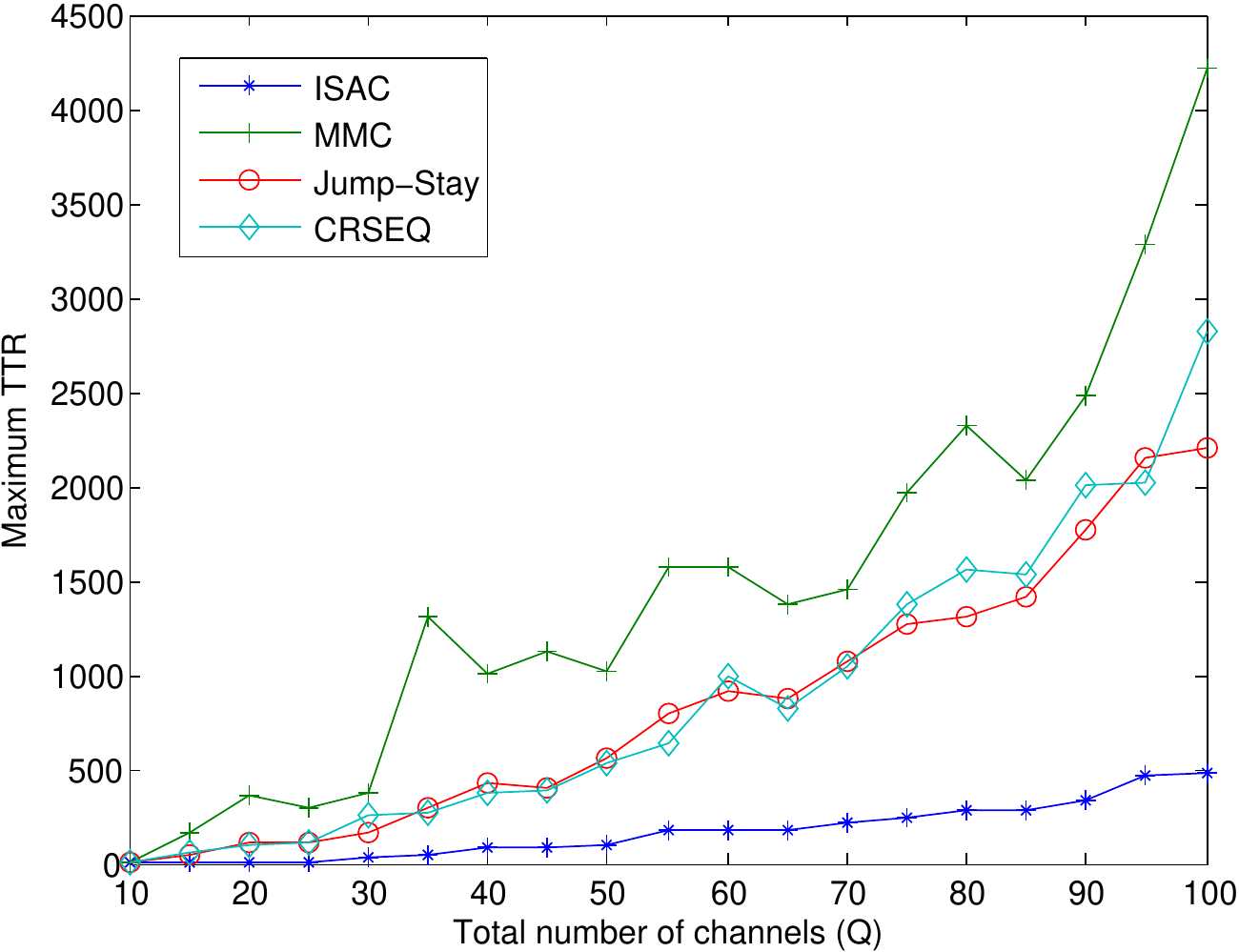}}
\hspace{1in}
\subfigure[Maximum TTR VS. $Q$ (without MMC)]{
\label{fig:subfig:b} %% label for second subfigure
\includegraphics[scale=0.5]{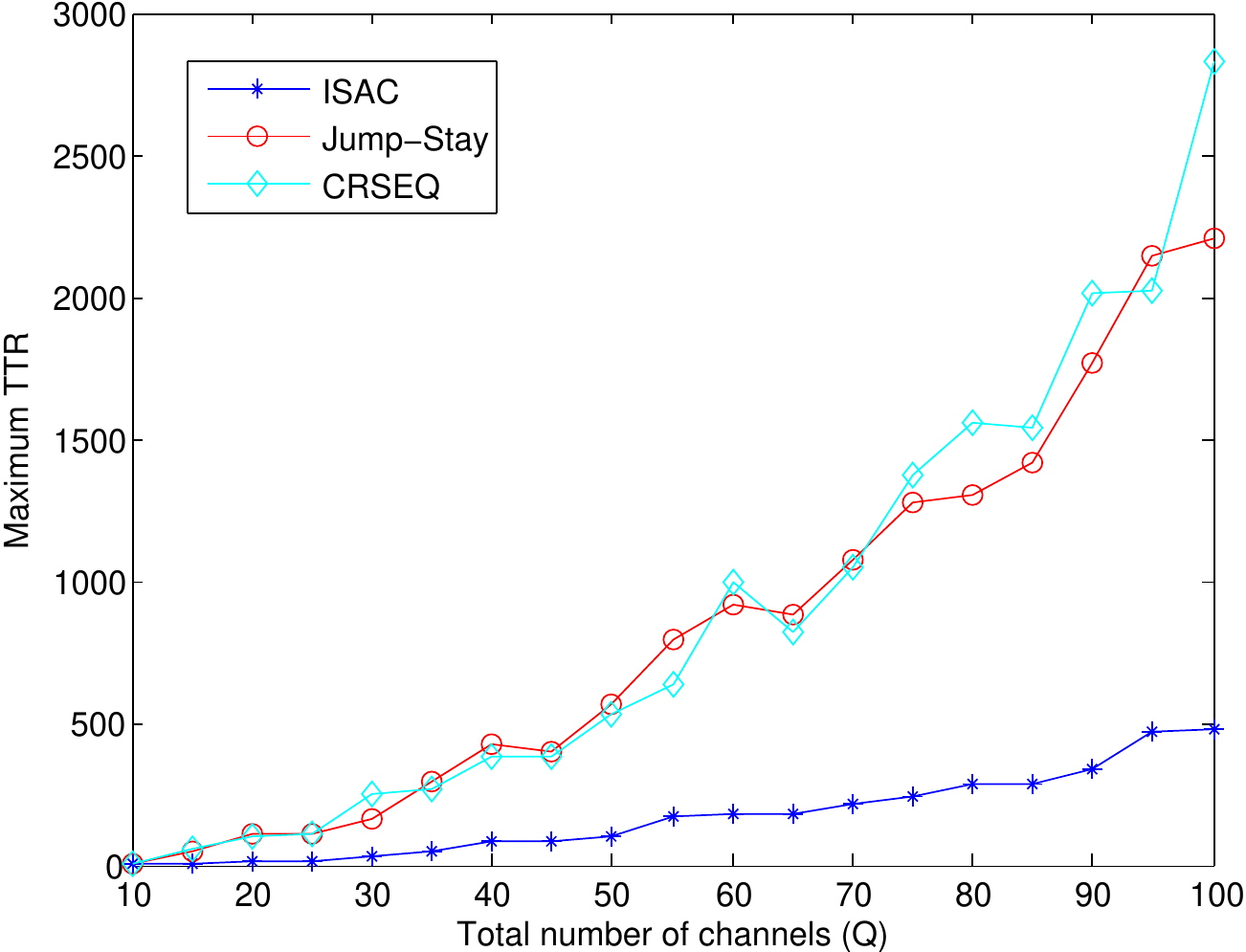}}
\hspace{1in}
\subfigure[Variance of TTR VS. $Q$]{
\label{fig:subfig:b} %% label for second subfigure
\includegraphics[scale=0.5]{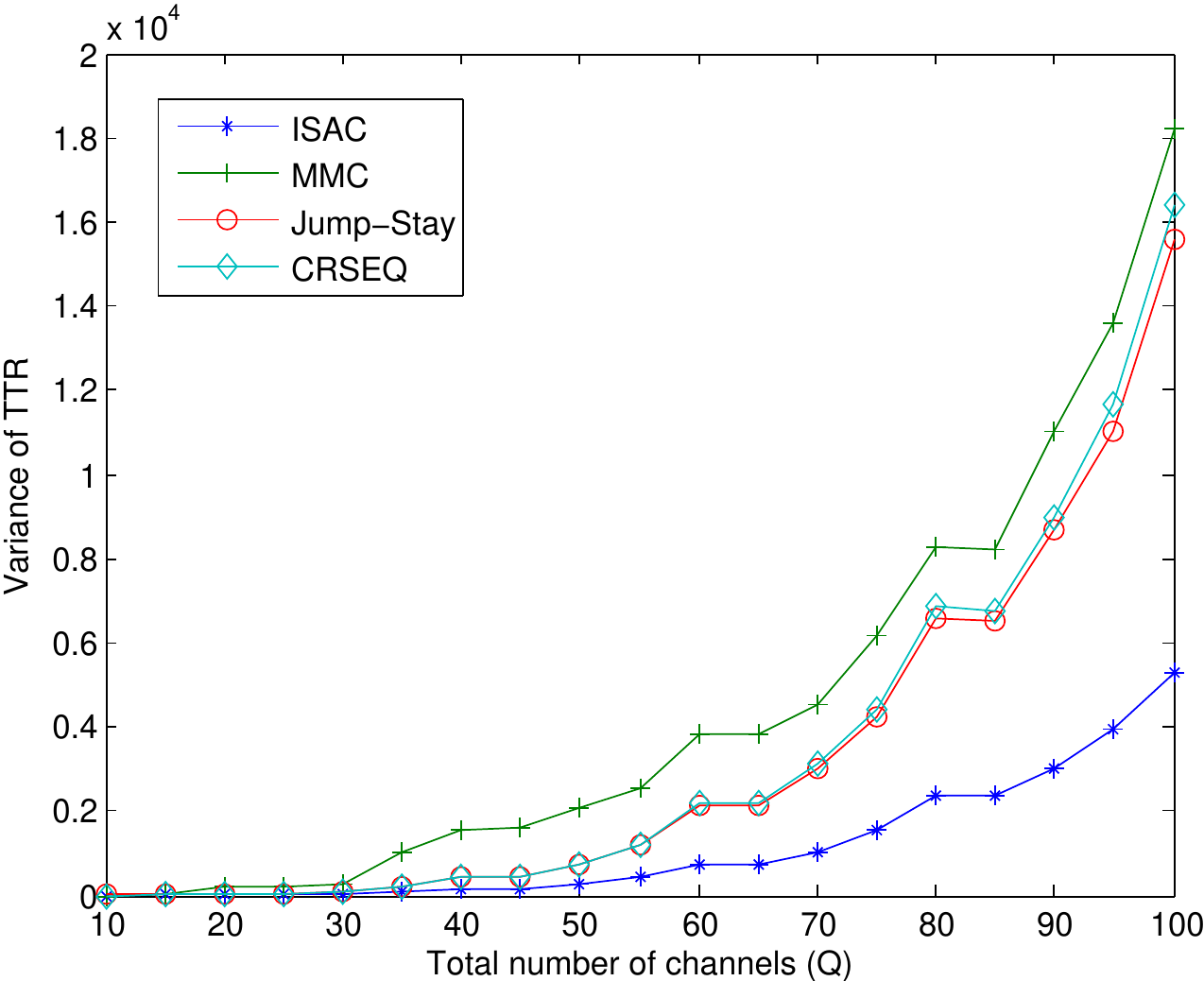}}
\caption{Comparison of different algorithms under the asymmetric model when $\theta=0.1$ and $G=1$.}
%\vspace{-0.2in}
\end{figure}
\item \emph{Moderate $\theta$}: Then we study the scenario when $\theta$ is moderate, that is, there are a moderate part of channels are available to users. We set $\theta=0.4$, $G=0.25\theta Q$. That is, there are 40\% channels are available to the users and 25\% among them are commonly-available to both of the users. Fig. 11 shows the average TTRs, the maximum TTRs and the variances of TTR of different algorithms against $Q$.  There is no large gap between different algorithms in terms of average TTR. For example, when there are 50 channels, the average TTRs of ISAC, MMC, Jump-Stay and CRSEQ are 75.94, 90.37, 83.09 and 91.20, respectively. However, ISAC significantly outperforms all the existing algorithms in terms of the maximum TTR. When there are 50 channels, the maximum TTRs are 758, 3482, 1121 and 1203, respectively. Its maximum TTR is almost one fifth of MMC. In this example, $m\leq0.45\times50=22.5$, $G=0.1\times50=5$. According to Theorem IV.2, $MTTR\leq 2\times m_pn-2G+2=2\times 23\times 22-2\times5+2=1004$. This result shows that the upper bound given by Theorem IV.2 is not tight when $G$ is large under the asymmetric model. The variances of TTR of ISAC, MMC, Jump-Stay and CRSEQ in this case are 5175.16, 16475.76, 7156.27 and 9463.17, respectively. The performance of ISAC is the most stable one.
\begin{figure}
\centering
\subfigure[Average TTR VS. $Q$]{
\label{fig:subfig:a} %% label for first subfigure
\includegraphics[scale=0.5]{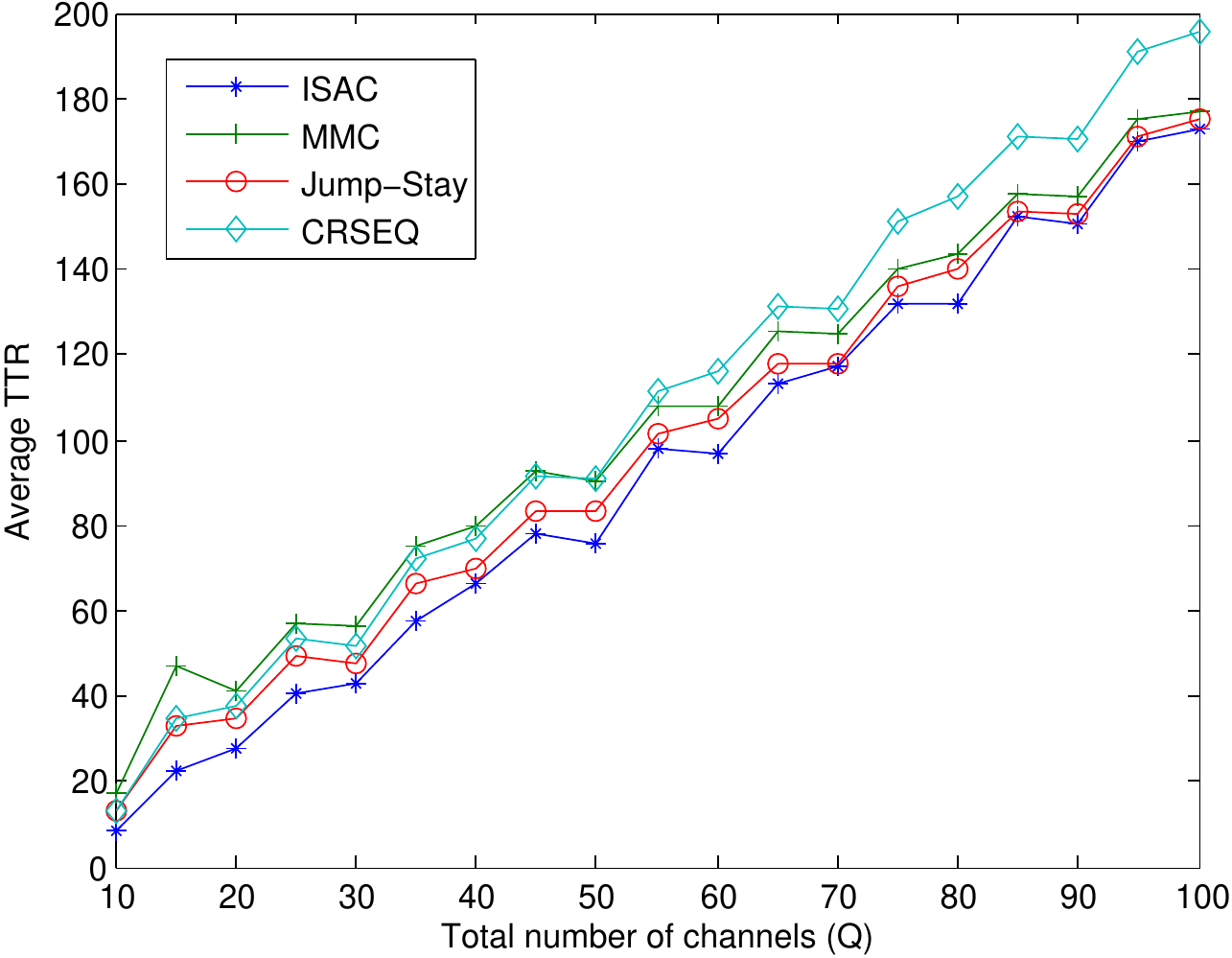}}
\hspace{1in}
\subfigure[Maximum TTR VS. $Q$]{
\label{fig:subfig:b} %% label for second subfigure
\includegraphics[scale=0.5]{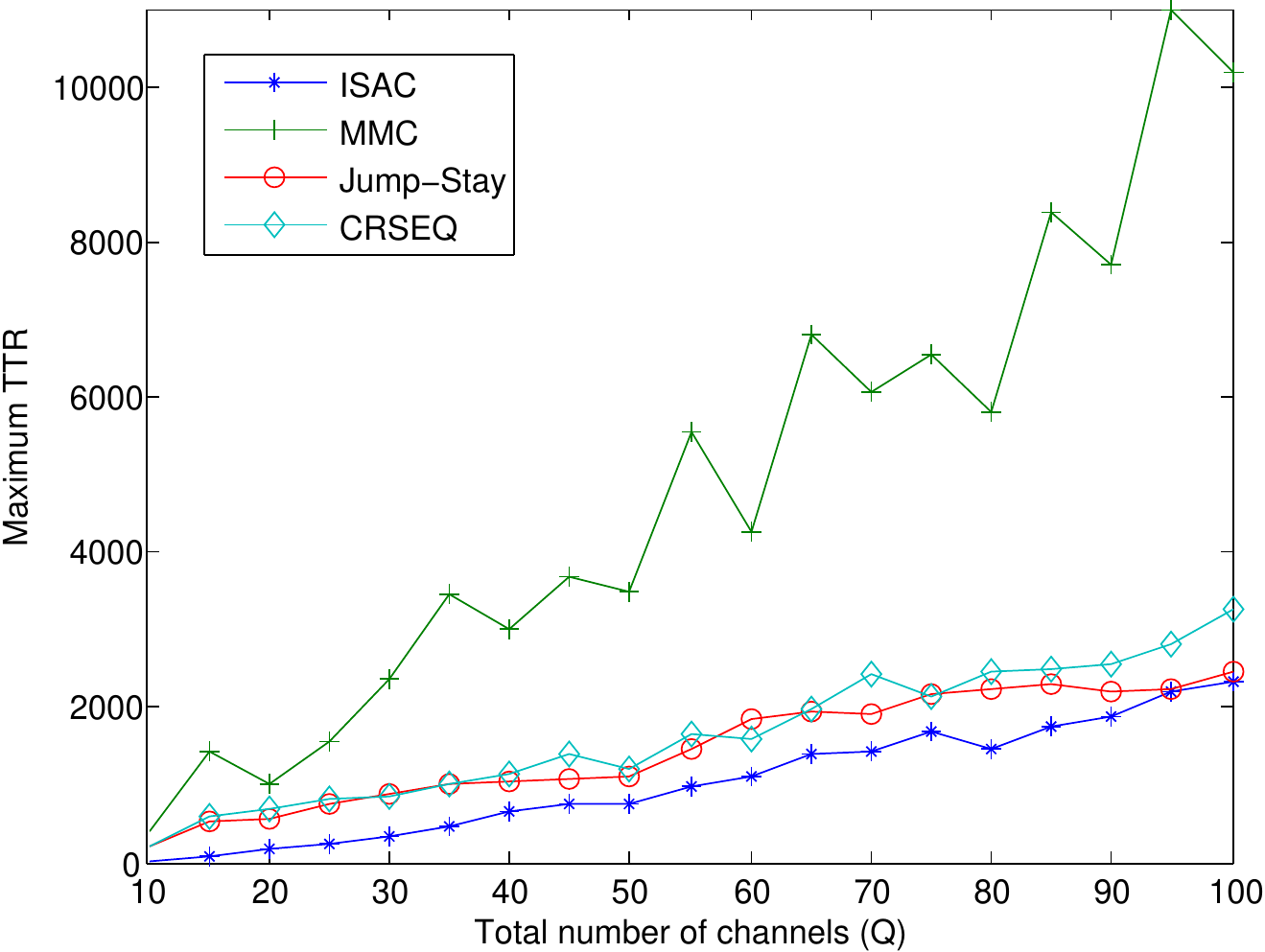}}
\hspace{1in}
\subfigure[Maximum TTR VS. $Q$ (without MMC)]{
\label{fig:subfig:b} %% label for second subfigure
\includegraphics[scale=0.5]{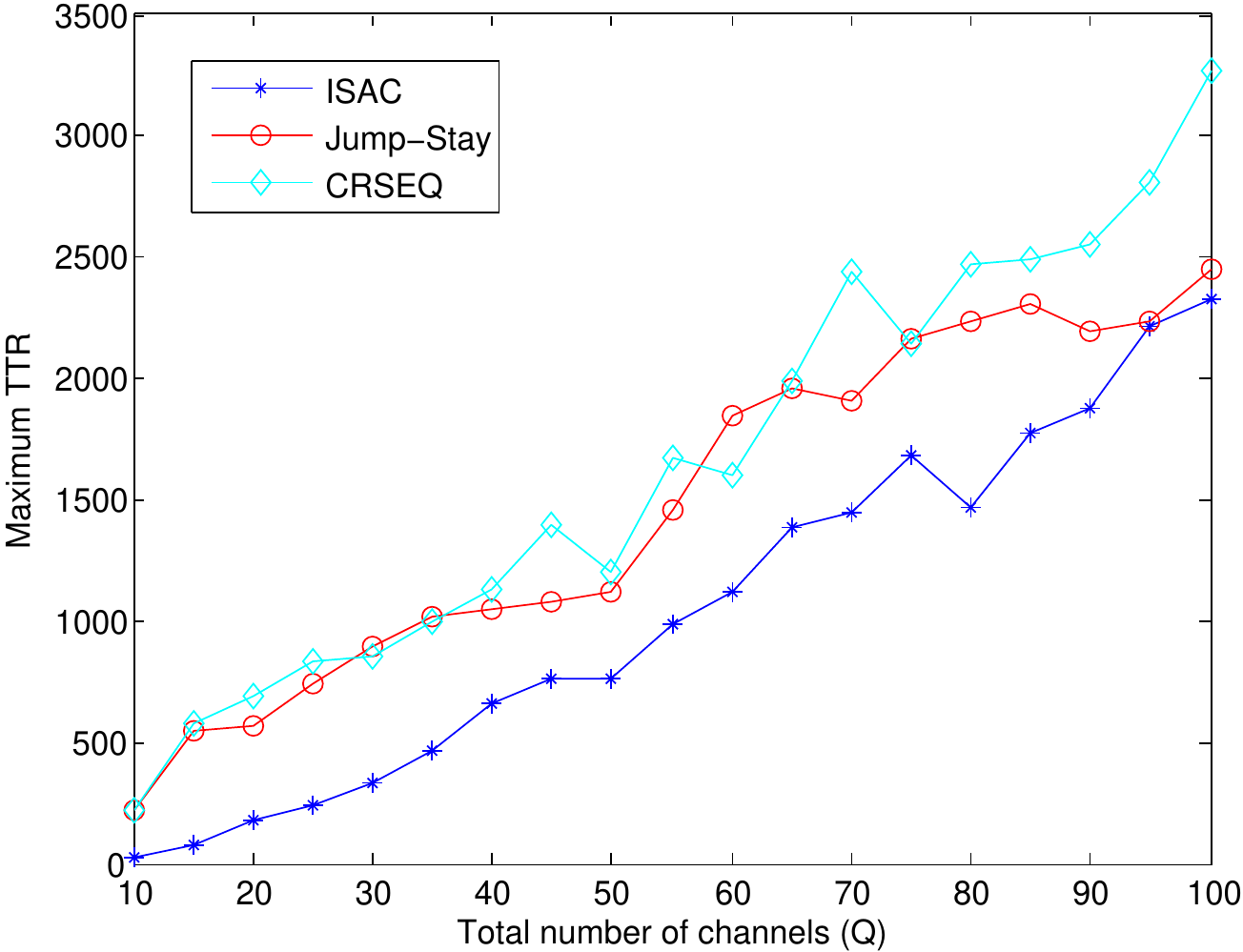}}
\hspace{1in}
\subfigure[Variance of TTR VS. $Q$]{
\label{fig:subfig:b} %% label for second subfigure
\includegraphics[scale=0.5]{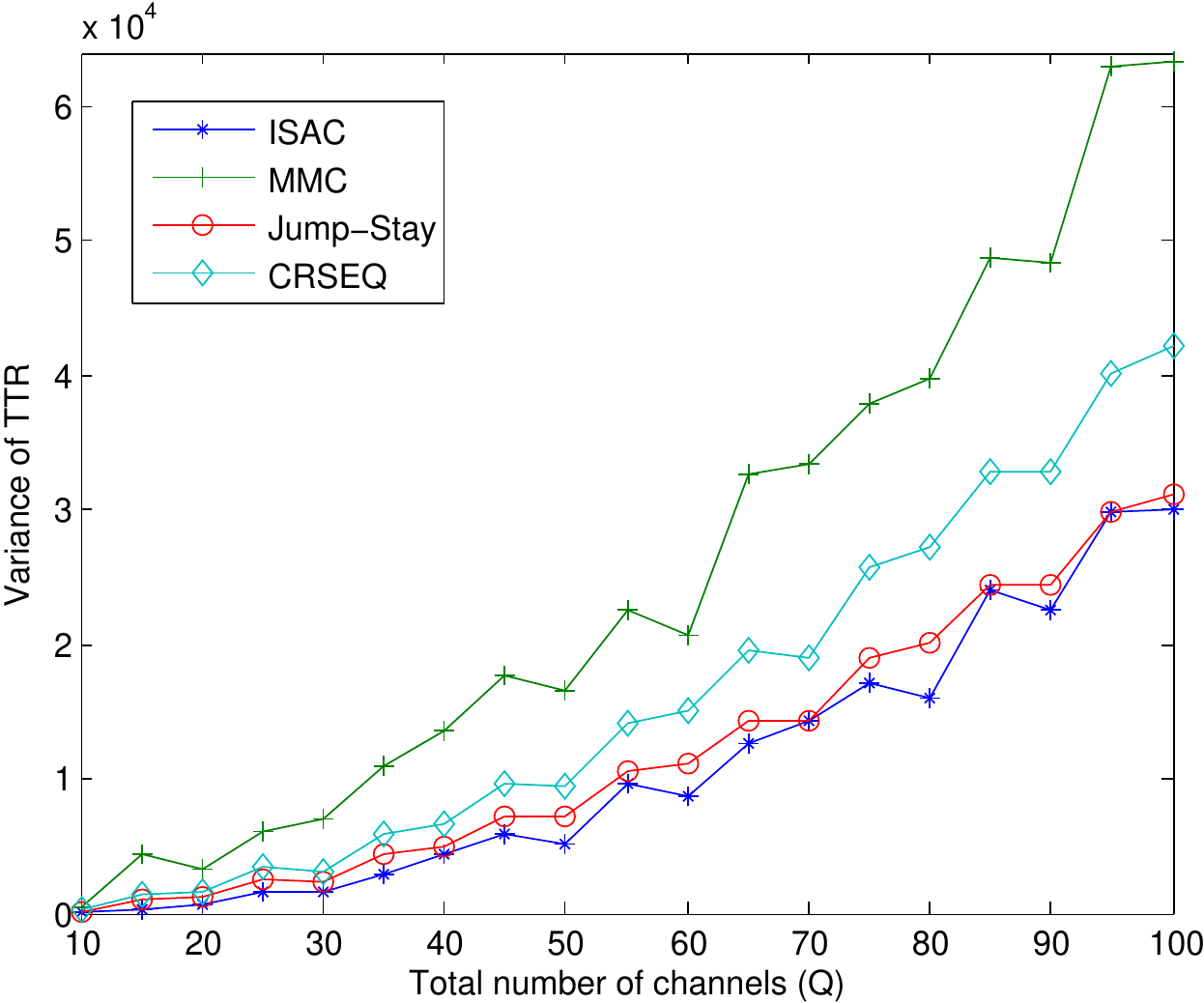}}
\caption{Comparison of different algorithms under the asymmetric model when $\theta=0.4$ and $G=0.25\theta Q$.}
%\vspace{-0.2in}
\end{figure}
\item \emph{Large $\theta$}: We now study the scenario when $\theta$ is large, that is, most channels are available to users. We set $\theta=0.8$, $G=0.75\theta Q$. That is, there are 80\% channels are available to the users and 75\% of them are commonly-available to both of the users. Fig. 12 shows the average TTRs, the maximum TTRs and the variances of TTR of different algorithms against $Q$.  The performance is similar to the scenario when $\theta=0.4$. For example, when there are 50 channels, the average TTRs of ISAC, MMC, Jump-Stay and CRSEQ are 59.53, 57.46, 53.76 and 87.60 while the maximum TTRs are 654, 10190, 993 and 1747, respectively. Maximum TTR of ISAC is almost one twentieth of MMC. The variances of TTR of ISAC, MMC, Jump-Stay and CRSEQ in this case are 2723.60, 11594.98, 2682.32 and 7363.49, respectively. The performance of ISAC and Jump-Stay are more stable than other two algorithms.
\begin{figure}
\centering
\subfigure[Average TTR VS. $Q$]{
\label{fig:subfig:a} %% label for first subfigure
\includegraphics[scale=0.5]{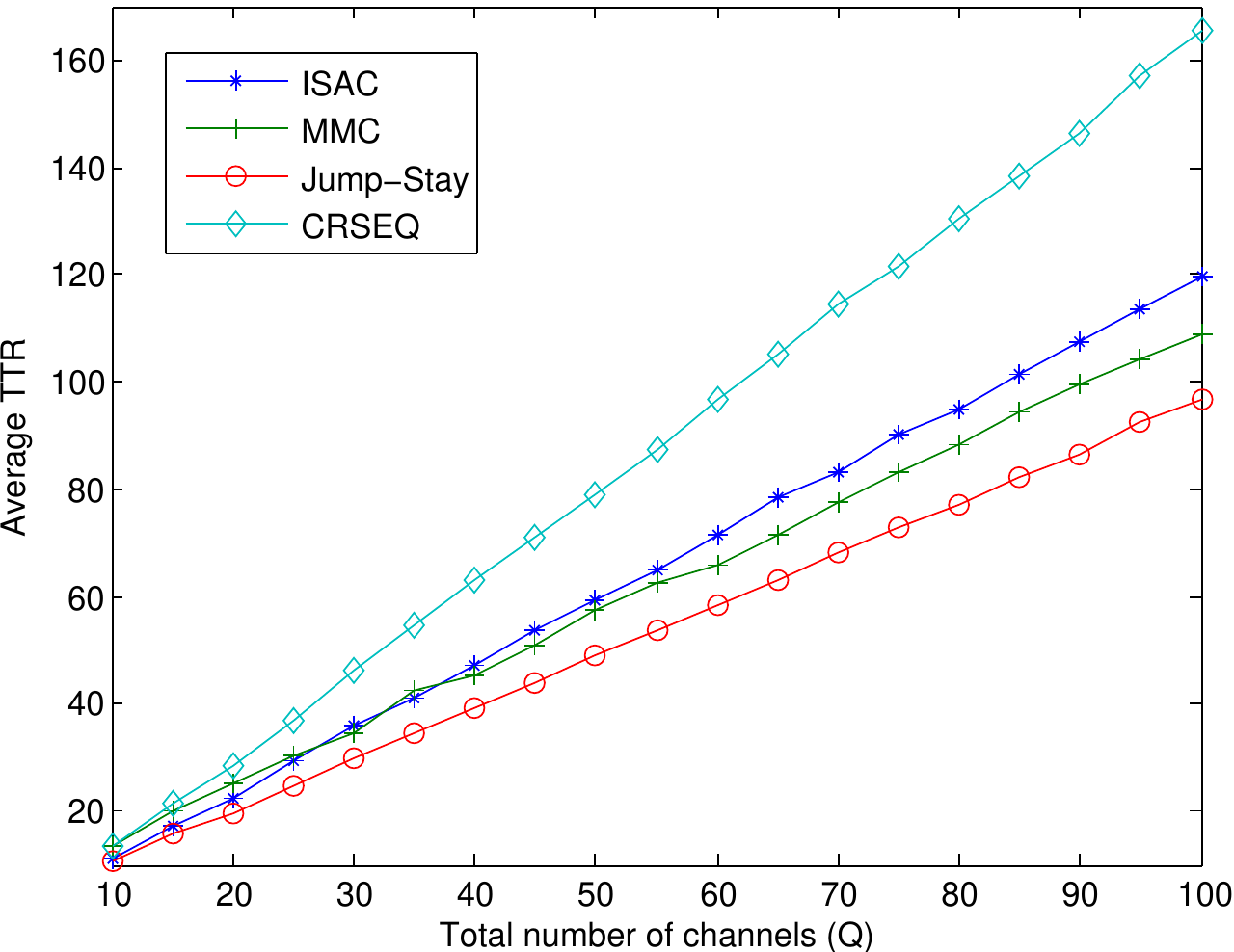}}
\hspace{1in}
\subfigure[Maximum TTR VS. $Q$]{
\label{fig:subfig:b} %% label for second subfigure
\includegraphics[scale=0.5]{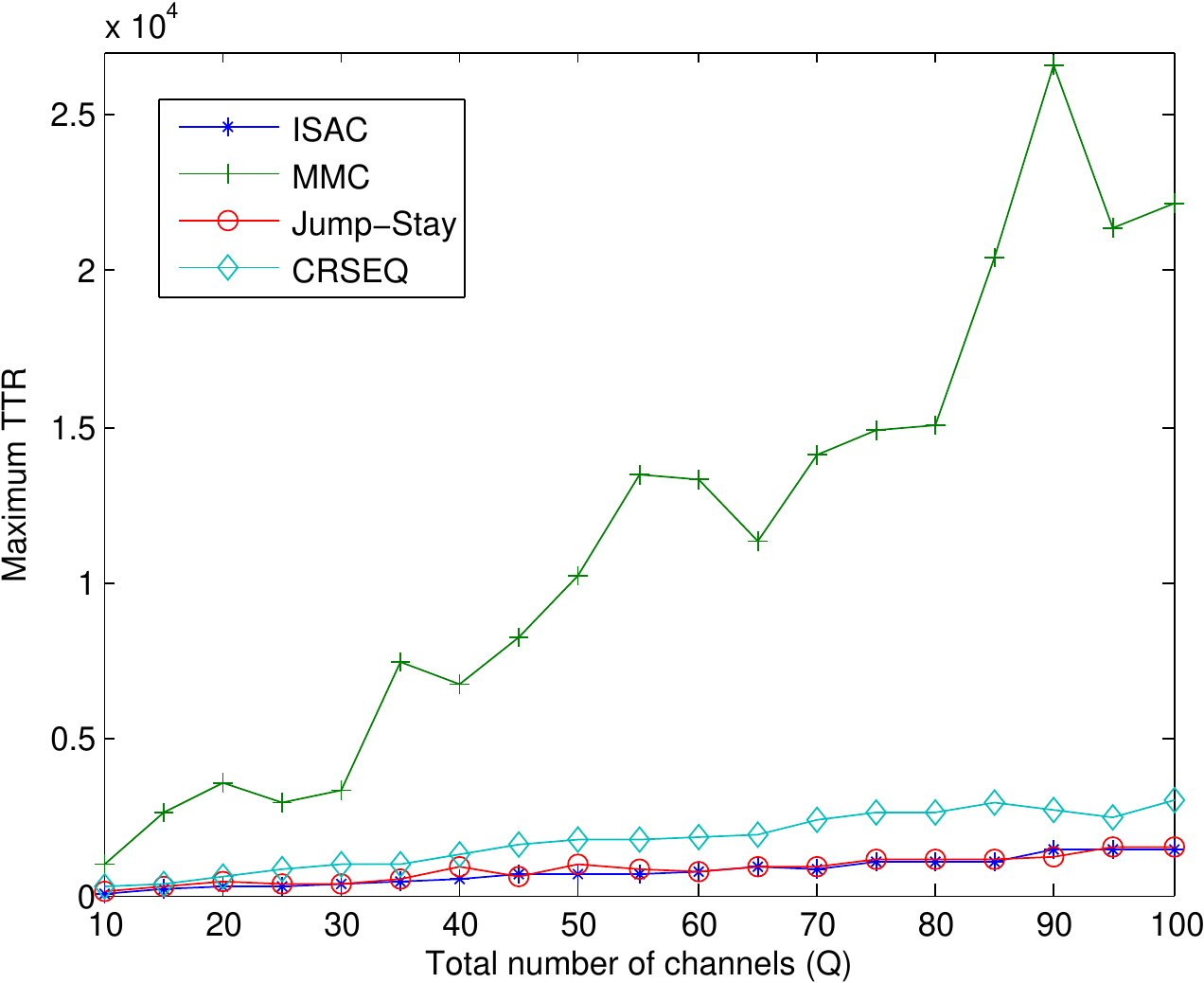}}
\hspace{1in}
\subfigure[Maximum TTR VS. $Q$ (without MMC)]{
\label{fig:subfig:b} %% label for second subfigure
\includegraphics[scale=0.5]{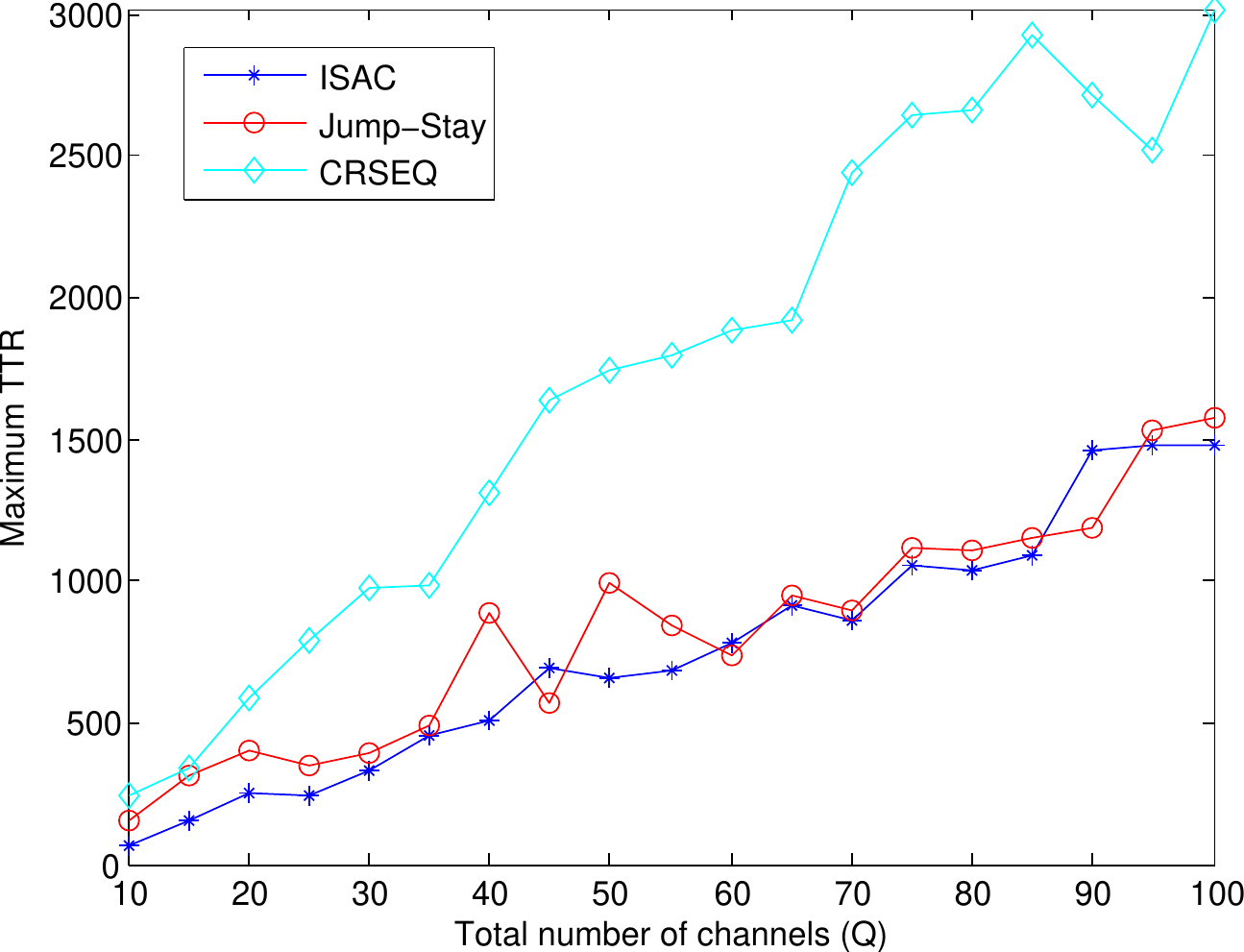}}
\hspace{1in}
\subfigure[Variance of TTR VS. $Q$]{
\label{fig:subfig:b} %% label for second subfigure
\includegraphics[scale=0.5]{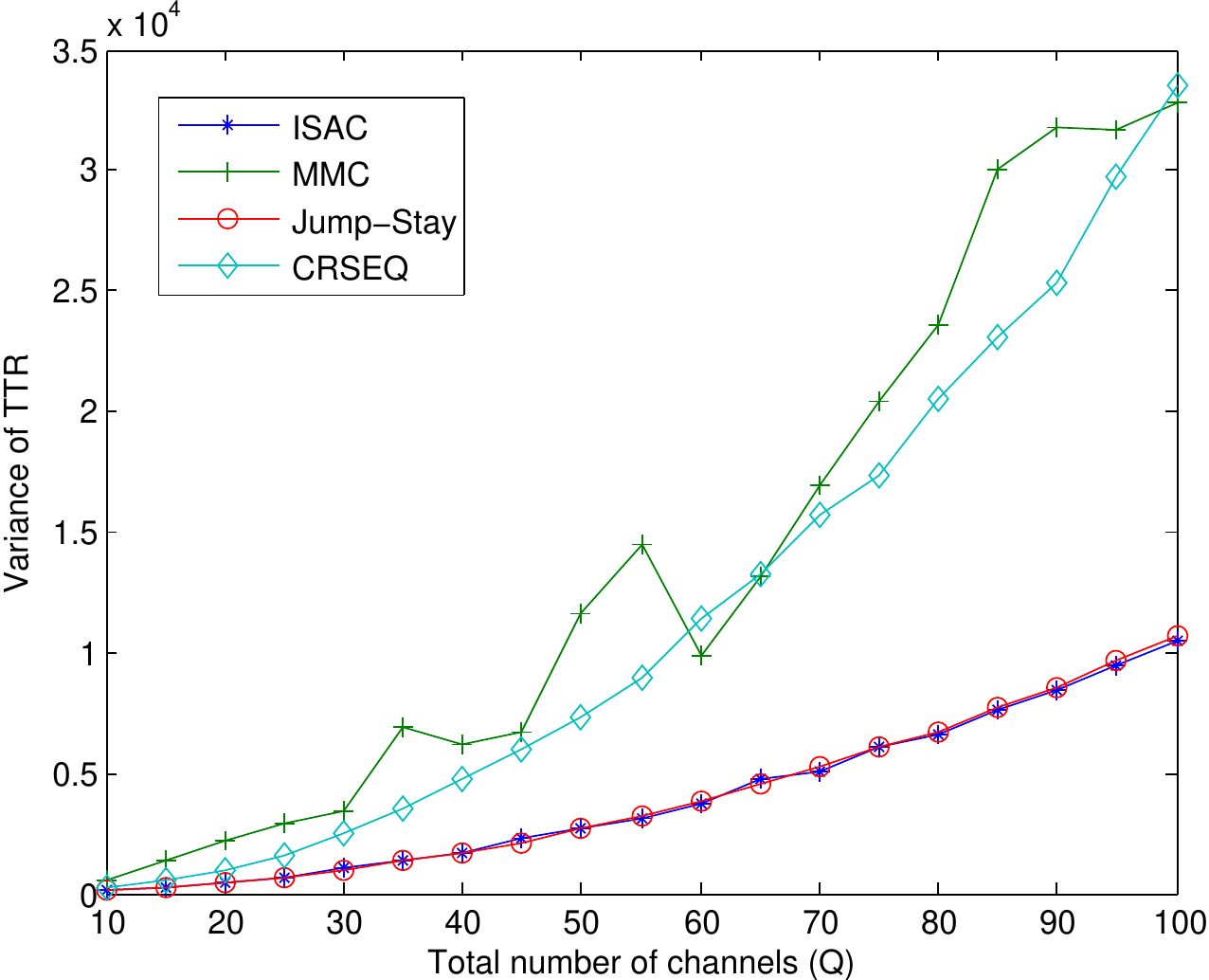}}
\caption{Comparison of different algorithms under the asymmetric model when $\theta=0.8$ and $G=0.75\theta Q$.}
%\vspace{-0.2in}
\end{figure}
\end{itemize}
\subsection{Influence of $m$ or $n$ for fixed $Q$}
Jump-Stay and CRSEQ generate CH sequence based on the whole channel set while Random and ISAC generate CH sequence based on the available channel set only. When we fix the total number of channels $Q$, we will get the same average TTR and maximum TTR for different $(m, n)$ if we apply Jump-Stay or CRSEQ. However, we will get different results if we apply ISAC or MMC. Fig. 13 shows both the average TTRs and maximum TTRs of different algorithms against $m$ (or $n$) when we fix $Q=60$ and $G=m$. We adopt eight values which are $m=0.1Q, 0.2Q, ..., 0.8Q$. According to Fig. 13, all algorithms get very different results when the number of available channels increases but the number of all channels is fixed. This is because there is a random-replacement in Jump-Stay and CRSEQ. When the channel in CH sequence is not available to the user, it will randomly select an available channel to replace. The chance to achieve rendezvous successfully in such time slots will increase significantly when the number of available channels is large. In terms of our ISAC algorithm, the eight simulated results are 12, 24, 36, 57, 60, 72, 84 and 92, respectively. According to Theorem IV.1, the eight theoretical results are 13, 25, 37, 57, 61, 73, 85 and 105, respectively. These results show that the upper bound given by Theorem IV.1 is tight.
\begin{figure}
\centering
\subfigure[Average TTR VS. $m/Q$]{
\label{fig:subfig:a} %% label for first subfigure
\includegraphics[scale=0.5]{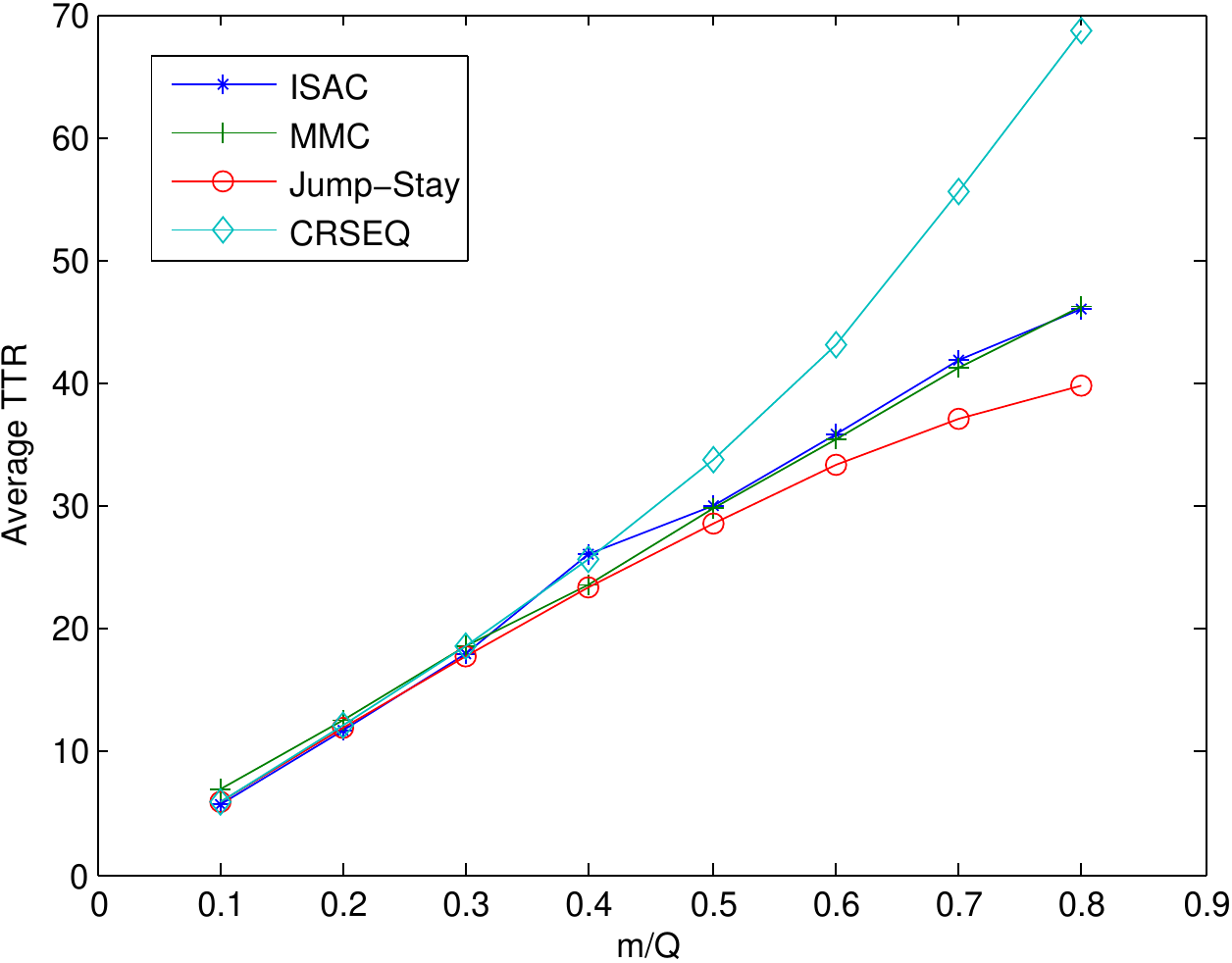}}
\hspace{1in}
\subfigure[Maximum TTR VS. $m/Q$]{
\label{fig:subfig:b} %% label for second subfigure
\includegraphics[scale=0.5]{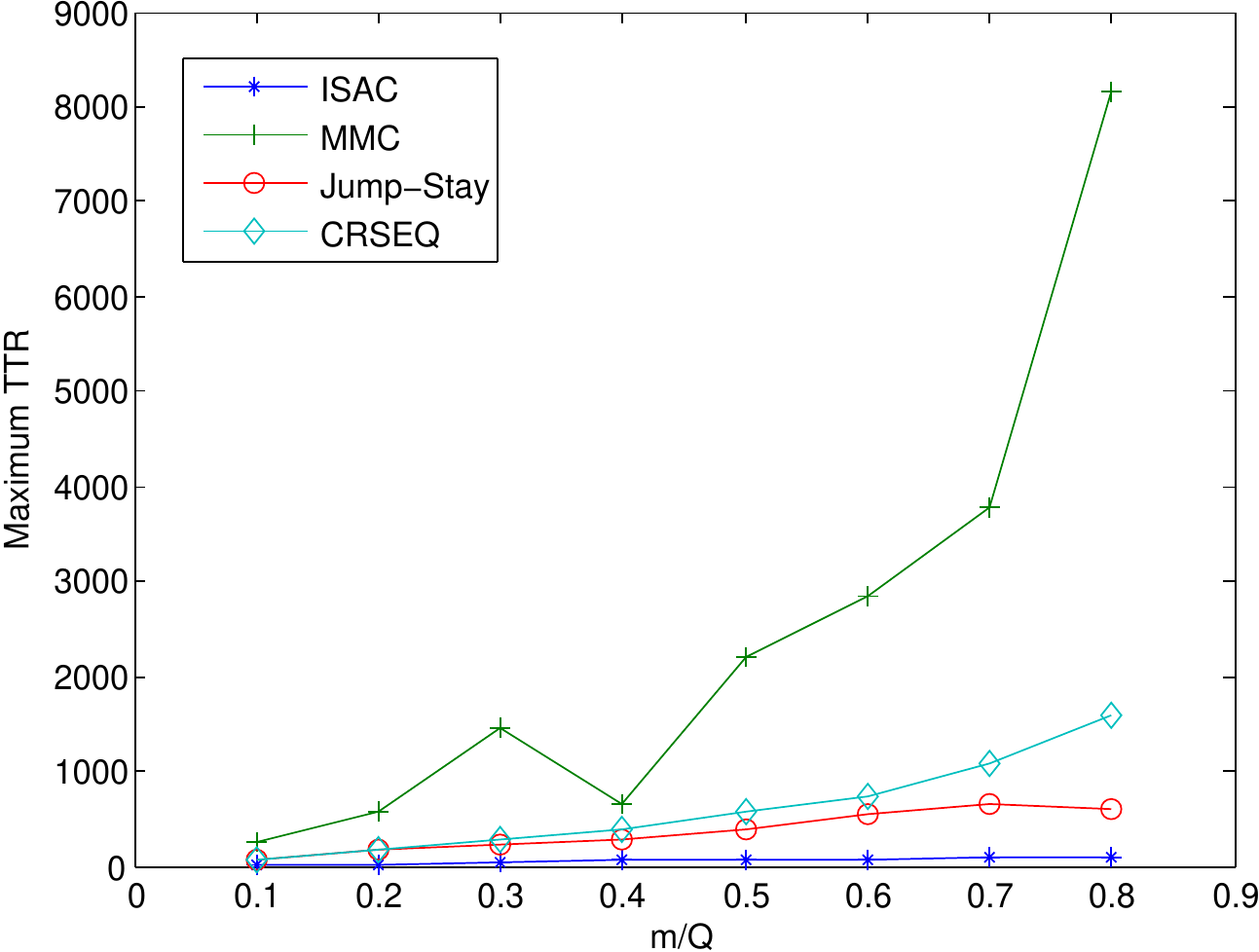}}
\hspace{1in}
\subfigure[Maximum TTR VS. $m/Q$ (without MMC)]{
\label{fig:subfig:b} %% label for second subfigure
\includegraphics[scale=0.5]{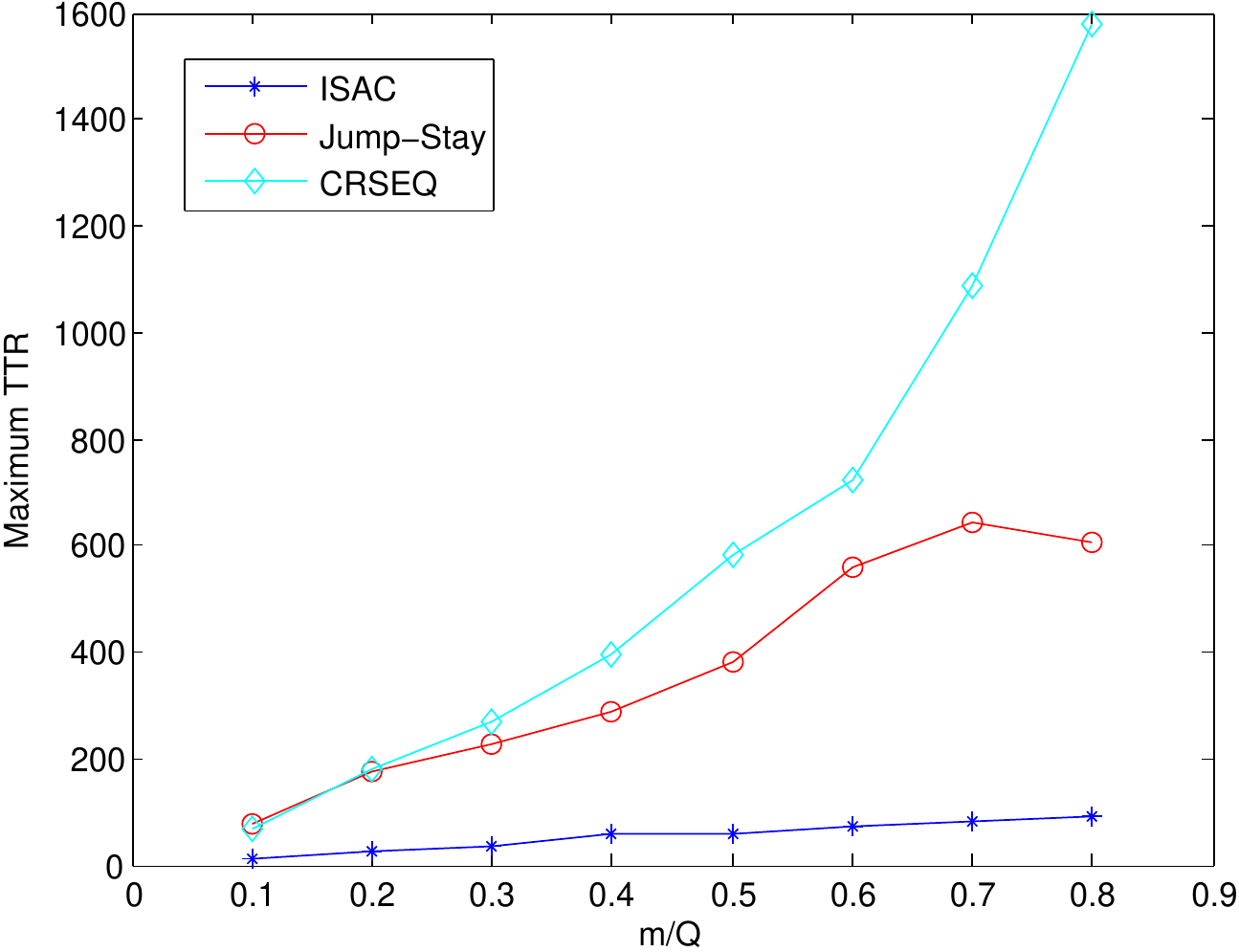}}
\hspace{1in}
\subfigure[Variance of TTR VS. $m/Q$]{
\label{fig:subfig:b} %% label for second subfigure
\includegraphics[scale=0.5]{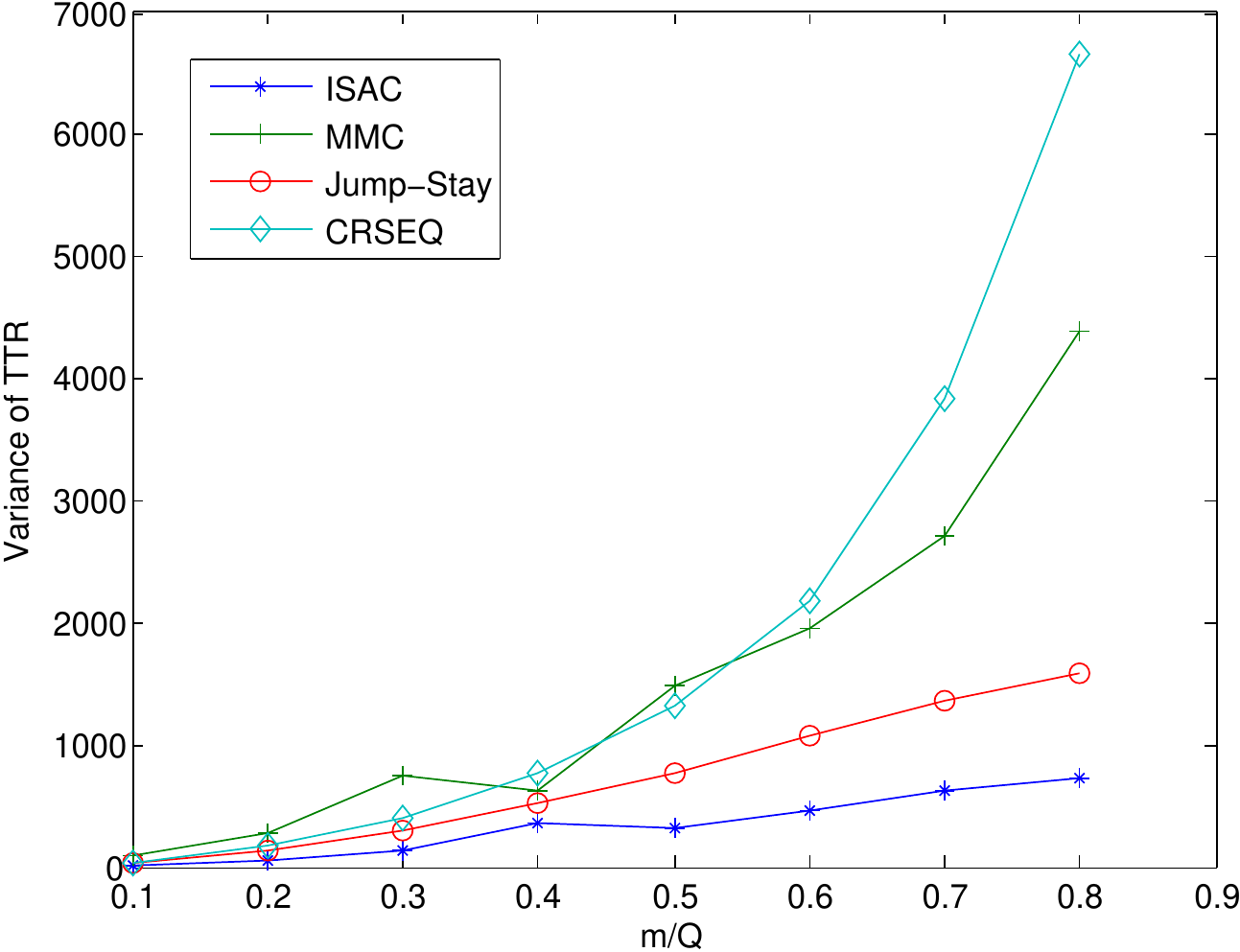}}
\caption{Influence of $m$ on rendezvous performance when $Q=60$.}
%\vspace{-0.2in}
\end{figure}
\subsection{Influence of $G$ for fixed $m$}
In this subsection, we study another basic property of the proposed approach: how does the number of commonly-available channels of the two users affect the rendezvous performance? We let $\theta=0.5$ and take five values of $G$ which are $G=0.2\theta Q$, $G=0.4\theta Q$, $G=0.6\theta Q$, $G=0.8\theta Q$, $G=\theta Q$. Fig. 14 shows the results when we apply the ISAC algorithm. When there are 60 channels, the average TTRs of five scenarios are 157.79, 84.76, 58.03, 42.82 and 29.91 while the maximum TTRs are 1747, 891, 659, 415 and 60, respectively. In this example, $Q=60$, $m\leq0.55\times60=33$, $m_p=37$. According to Theorem IV.1 and Theorem IV.2, however, the upper bound of MTTR in five scenarios are 2428, 2416, 2404, 2392 and 73, respectively. We note that our upper bound under the symmetric model is tight while the upper bound under the asymmetric model is less tight especially when $G$ is large.
\begin{figure}
\centering
\subfigure[Average TTR VS. $Q$]{
\label{fig:subfig:a} %% label for first subfigure
\includegraphics[scale=0.5]{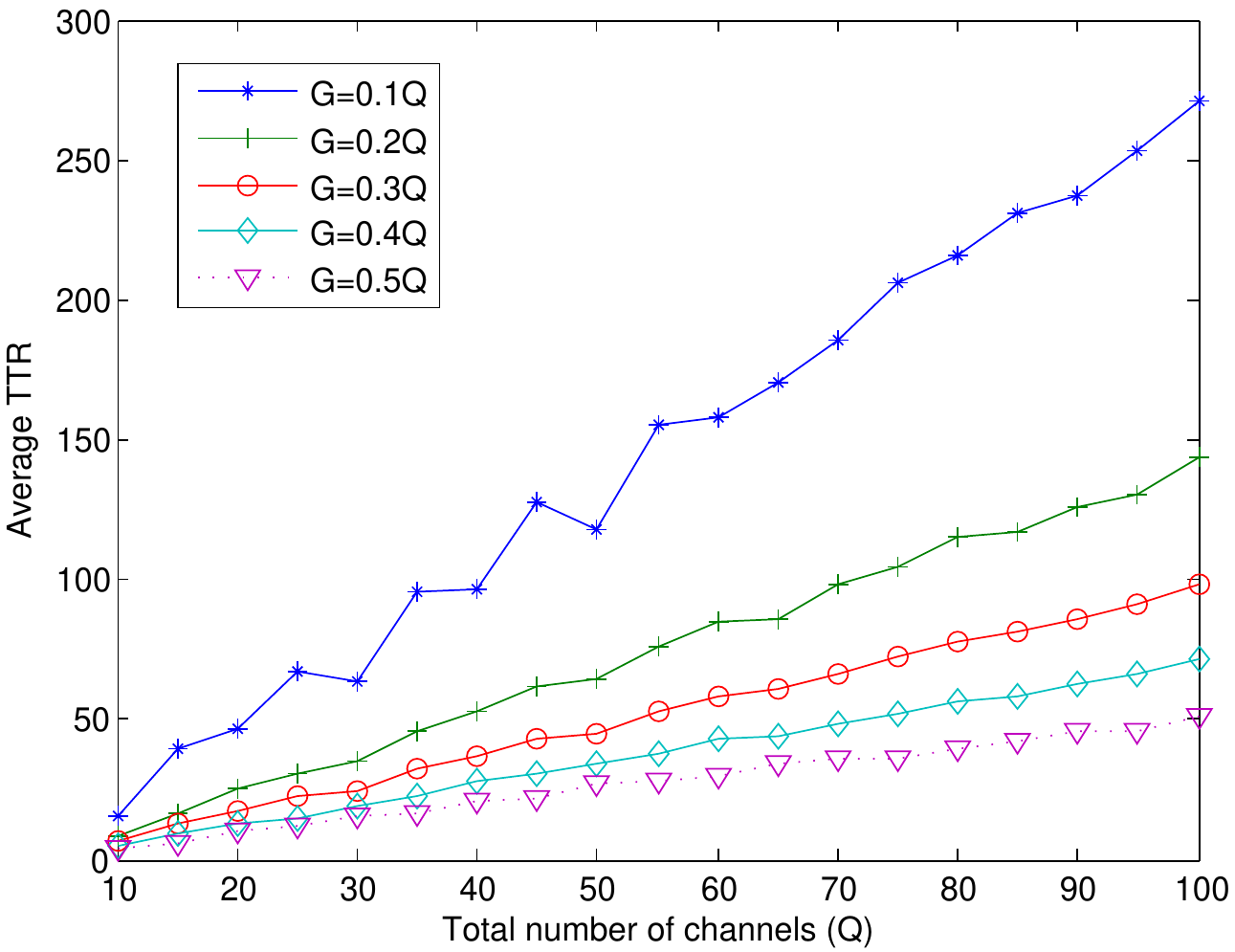}}
\hspace{1in}
\subfigure[Maximum TTR VS. $Q$]{
\label{fig:subfig:b} %% label for second subfigure
\includegraphics[scale=0.5]{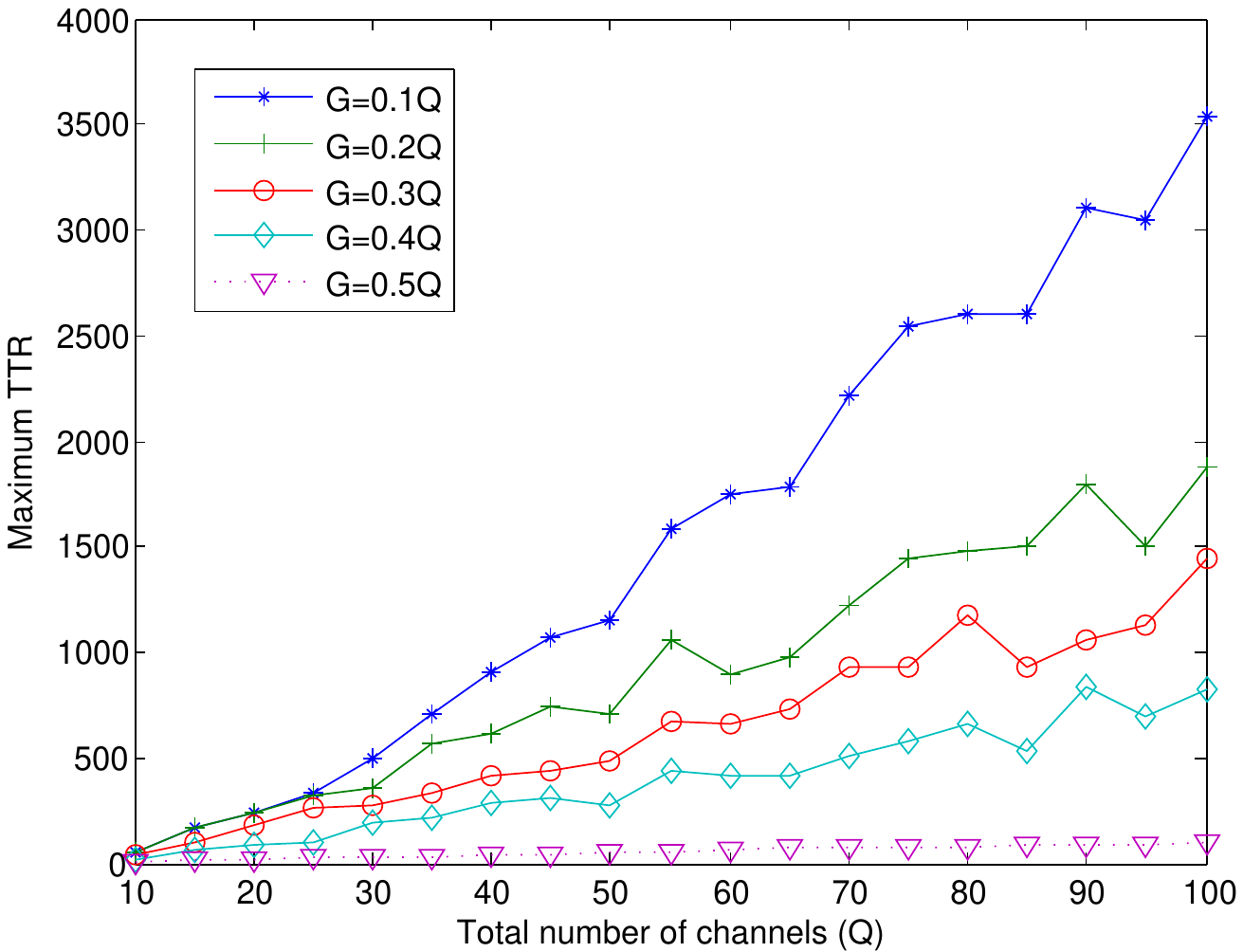}}
\caption{Influence of $G$ on rendezvous performance when $\theta=0.5$.}
%\vspace{-0.2in}
\end{figure}
\section{Conclusions}
We designed a new rendezvous algorithm, called ISAC, for cognitive radio networks. ISAC is the first one in the literature that generates CH sequences based on the available channel set instead of the whole channel set (the available channel set is subset of the whole channel set) while providing guaranteed rendezvous. We proved that ISAC provides guaranteed rendezvous and derived upper bounds on the maximum TTR (MTTR) under both the symmetric model and the asymmetric model. These upper bounds are expressed in terms of the number of available channels instead of the total number of channels ($2m_p-1$ under the symmetric model and $2m_pn-2G+2$ under the asymmetric model ), and  they are tighter than the upper bounds of the existing rendezvous algorithms. We conducted extensive computer simulation for performance evaluation and observed the following properties:
\begin{itemize}
\item In terms of the maximum TTR, ISAC significantly outperforms other algorithms in almost all scenarios. For example, ISAC reduces the maximum TTR up to 97\% when $Q=100$ and $G=10$ in Fig. 7(a). ISAC is suitable to many QoS-concern applications of CRNs which desire small MTTR.
\item In terms of the average TTR, performance of ISAC is better than those of other algorithms when the available channel set is a small subset of the whole channel set (e.g., it reduces the average TTR up to 45.8\% in Fig. 10(a)).
\item In terms of the variance of TTR, ISAC is the most stable algorithm among all the rendezvous algorithms in our simulation.
\item The upper bound on MTTR under the symmetric model is tight while the upper bound under the asymmetric model is less tight especially when $G$ is large (e.g., greater than $0.2\theta Q$ in Fig. 14).
\end{itemize}


\begin{thebibliography}{1}

\bibitem{IEEEhowto:kopka}
%This is an example of a book reference
I. Akyildiz, W. Lee, M. Vuran, and S. Mohanty, ``NeXt Generation Dynamic Spectrum Access Cognitive Radio Wireless Networks: A Survey," \emph{Computer Networks}, vol. 50, no. 13, pp. 2127-2159, 2006.
\bibitem{IEEEhowto:kopka}
T. Ycek and H. Arslan, ``A Survey of Spectrum Sensing Algorithms for Cognitive Radio Applications," \emph{IEEE Communications Surveys \& Tutorials}, vol. 11, no. 1, pp. 116-130, 2009.
\bibitem{IEEEhowto:kopka}
P. Wang, L. Xiao, S. Zhou, and J. Wang, ``Optimization of Detection Time for Channel Efficiency in Cognitive Radio Systems," \emph{Proc. of IEEE WCNC}, pp.111-115, Mar. 2007.
\bibitem{IEEEhowto:kopka}
H. Liu, Z. Lin, X. Chu, and Y. W. Leung, ``Taxonomy and Challenges of Rendezvous Algorithms in Cognitive Radio Networks," invited position paper, \emph{Proc. of IEEE ICNC}, pp. 645-649, Hawaii, 2012.
\bibitem{IEEEhowto:kopka}
H. Liu,  Z. Lin,  X. Chu, and Y. W. Leung, ``Jump-Stay Rendezvous Algorithm for Cognitive Radio Networks," \emph{IEEE Trans. Parallel and Distributed  Systems}, vol. 23, no. 10, pp. 1867-1881, 2012.
\bibitem{IEEEhowto:kopka}
Z. Lin, H. Liu, X. Chu, and Y.-W. Leung, ``Jump-stay Based Channel-hopping Algorithm With Guaranteed Rendezvous for Cognitive Radio Networks," \emph{Proc. of IEEE INFOCOM}, pp. 2444-2452, Apr. 2011.
\bibitem{IEEEhowto:kopka}
Z. Lin,  H. Liu,  X. Chu, and Y. W. Leung, ``Enhanced Jump-Stay Rendezvous Algorithm for Cognitive Radio Networks," \emph{IEEE Communications Letters}, vol. 17, no. 9, pp. 1742-1745, 2013.
\bibitem{IEEEhowto:kopka}
J. Shin, D. Yang, and C. Kim, ``A Channel Rendezvous Scheme for Cognitive Radio Networks," \emph{IEEE Communications Letters}, vol. 14, no. 10, pp. 954-956, 2010.
\bibitem{IEEEhowto:kopka}
N. C. Theis, R. W. Thomas, and L. A. DaSilva, ``Rendezvous for cognitive radios," \emph{IEEE Transactions on Mobile Computing}, vol. 10, no. 2, pp. 216-227, 2010.
\bibitem{IEEEhowto:kopka}
K. Bian, J.-M. Park, and R. Chen, ``Control Channel Establishment in Cognitive Radio Networks Using Channel Hopping," \emph{IEEE Journal on Selected Areas in Communications}, vol. 29, no. 4, pp. 689-703, 2011.
\bibitem{IEEEhowto:kopka}
M. M. Buddhikot, P. Kolodzy, S. Miller, K. Ryan, and J. Evans, ``DIMSUMnet: New Directions in Wireless Networking Using Coordinated Dynamic Spectrum," \emph{Proc. of IEEE WoWMoM}, pp. 78-85, Jun. 2005.
\bibitem{IEEEhowto:kopka}
B. Horine and D. Turgut, ``Performance Analysis of Link Rendezvous Protocol for Cognitive Radio Networks," \emph{Proc. of CROWNCOM}, pp. 503-507, Aug. 2007.
\bibitem{IEEEhowto:kopka}
P. Sutton, K. Nolan, and L. Doyle, ``Cyclostationary Signatures in Practical Cognitive Radio Applications," \emph{IEEE Journal on Selected Areas in Communications}, vol. 26, no. 1, pp. 13-24, Jan. 2008.
\bibitem{IEEEhowto:kopka}
V. Brik, E. Rozner, S. Banerjee, and P. Bahl, ``DSAP: A Protocol for Coordinated Spectrum Access," \emph{Proc. of IEEE DySPAN}, pp. 611-614, Nov. 2005.
\bibitem{IEEEhowto:kopka}
J. Jia, Q. Zhang, and X. Shen, ``HC-MAC: A Hardware-constrained Cognitive MAC for Efficient Spectrum Management," \emph{IEEE Journal on Selected Areas in Communications}, vol. 26, no. 1, pp. 106-117, 2008.
\bibitem{IEEEhowto:kopka}
L. Ma, X. Han, and C.-C. Shen, ``Dynamic Open Spectrum Sharing for Wireless Ad Hoc Networks," \emph{Proc. of IEEE DySPAN}, pp. 203-213, Nov. 2005.
\bibitem{IEEEhowto:kopka}
J. Zhao, H. Zheng, and G.-H. Yang, ``Distributed Coordination in Dynamic Spectrum Allocation Networks," \emph{Proc. of IEEE DySPAN}, pp. 259-268, Nov. 2005.
\bibitem{IEEEhowto:kopka}
L. Lazos, S. Liu, and M. Krunz, ``Spectrum Opportunity-based Control Channel Assignment in Cognitive Radio Networks," \emph{Proc. of IEEE SECON}, pp. 1-9, Jun. 2009.
\bibitem{IEEEhowto:kopka}
K. Bian, J.-M. Park, and R. Chen, ``A Quorum-based Framework for Establishing Control Channels in Dynamic Spectrum Access Networks," \emph{Proc. of ACM MobiCom}, pp. 25-36, Sept. 2009.
\bibitem{IEEEhowto:kopka}
Y. Zhang et al., ``ETCH: Efficient Channel Hopping for Communication Rendezvous in Dynamic Spectrum Access Networks," \emph{Proc. of IEEE INFOCOM}, pp. 2471-2479, Apr. 2011.
\bibitem{IEEEhowto:kopka}
P. Bahl, R. Chandra, and J. Dunagan, ``SSCH: Slotted Seeded Channel Hopping for Capacity Improvement in IEEE 802.11 Ad-hoc Wireless Networks," \emph{Proc. of ACM MobiCom}, pp. 216-230, Sep. 2004.
\bibitem{IEEEhowto:kopka}
S. Krishnamurthy, M. Thoppian, S. Kuppa, R. Chandrasekaran, N. Mittal, S. Venkatesan, and R. Prakash, ``Time-efficient Distributed Layer-2 Auto-configuration for Cognitive Radio Networks," \emph{Computer Networks}, vol. 52, no. 4, pp. 831-849, 2008.
\bibitem{IEEEhowto:kopka}
H. Liu,  Z. Lin,  X. Chu, and Y. W. Leung, ``Ring-Walk Based Channel-Hopping Algorithm with Guaranteed Rendezvous for Cognitive Radio Networks," \emph{Proc. of IEEE/ACM International Conference on Green Computing and Communications}, pp. 755-760, Dec. 2010.
\bibitem{IEEEhowto:kopka}
D. Yang, J. Shin, and C. Kim, ``Deterministic Rendezvous Scheme in Multichannel Access Networks," \emph{Electronics Letters}, vol. 46, no. 20, pp. 1402-1404, 2010.
\bibitem{IEEEhowto:kopka}
K. Bian and J.-M. Park, ``Maximizing Rendezvous Diversity in Rendezvous Protocols for Decentralized Cognitive Radio Networks," \emph{IEEE Transactions on Mobile Computing}, vol. 12, no. 7, pp. 1294-1307, 2013.
\bibitem{IEEEhowto:kopka}
Z. Gu, Q.-S. Hua, Y. Wang, F. Lau, ``Nearly Optimal Asynchronous Blind Rendezvous Algorithm for Cognitive Radio Networks," \emph{Proc. of IEEE SECON}, pp. 371-379, Jun. 2013.
\bibitem{IEEEhowto:kopka}
C. Cormio and K. R. Chowdhury, ``Common Control Channel Design for Cognitive Radio Wireless Ad Hoc Networks Using Adaptive Frequency Hopping," \emph{Ad Hoc Networks}, vol. 8, pp. 430-438, 2010.
\bibitem{IEEEhowto:kopka}
C. Cordeiro and K. Challapali, ``C-MAC: A Cognitive MAC Protocol for Multi-Channel Wireless Networks," \emph{Proc. of IEEE DySPAN}, pp. 147-157, Apr. 2007.
\bibitem{IEEEhowto:kopka}
S. Romaszko, ``A Rendezvous Protocol with the Heterogeneous Spectrum Availability Analysis for Cognitive Radio Ad Hoc Networks," \emph{Journal of Electrical and Computer Engineering}, Article ID 715816. 2013.
\bibitem{IEEEhowto:kopka}
D. E. Knuth, \emph{The Art of Computer Programming}, vol. 1: Fundamental Algorithms, 3rd ed., Addison Wesley, 1997.
%This is an example of a Transactions article reference
%D.S. Coming and O.G. Staadt, "Velocity-Aligned Discrete Oriented Polytopes for Dynamic Collision Detection," IEEE Trans. Visualization and Computer Graphics, vol."1¤714,"1¤7 no."1¤71,"1¤7 pp. 1-12,"1¤7 Jan/Feb"1¤7 2008, doi:10.1109/TVCG.2007.70405.

%This is an example of a article from a conference proceeding
%H. Goto, Y. Hasegawa, and M. Tanaka, "Efficient Scheduling Focusing on the Duality of MPL Representation," Proc. IEEE Symp. Computational Intelligence in Scheduling (SCIS '07), pp. 57-64, Apr. 2007, doi:10.1109/SCIS.2007.367670.

%This is an example of a PrePrint reference
%J.M.P. Martinez, R.B. Llavori, M.J.A. Cabo, and T.B. Pedersen, "Integrating Data Warehouses with Web Data: A Survey," IEEE Trans. Knowledge and Data Eng., preprint, 21 Dec. 2007, doi:10.1109/TKDE.2007.190746.

%Again, see the IEEEtrans_HOWTO.pdf for several more bibliographical examples. Also, more style examples
%can be seen at http://www.computer.org/author/style/transref.htm
\end{thebibliography}
\end{document}